\documentstyle[aas2pp4,amssym,epsfig]{article}

\righthead{12\,$\mu$m Seyferts}
\lefthead{Hunt, L.K. et al.}

\begin{document}

\title{Morphology of the 12\,$\mu$m Seyfert Galaxies:
II. Optical and Near-Infrared Image Atlas\footnote{Based
on observations collected at the Calar Alto Observatory, the Cerro 
Tololo Interamerican Observatory, the European Southern Obsevatory, 
the Gornergrat Infrared Telescope,
the Lick Observatory, and the Nordic Optical Telescope.
}
} 

\author{ L. K. Hunt}
\affil{ C. A. I. S. M. I. - C. N. R. \\
Largo E. Fermi 5, I-50125 Firenze, Italy\\
Electronic mail: hunt@arcetri.astro.it}

\author { M. A. Malkan}
\affil{ University of California \\
Department of Astronomy, 405 Hilgard Ave., \\ 
Los Angeles, CA, U.S.A. 90095-1562 \\
Electronic mail: malkan@bonnie.astro.ucla.edu}

\author{B. Rush}
\affil{ University of California \\
Department of Astronomy, 405 Hilgard Ave., \\ 
Los Angeles, CA, U.S.A. 90095-1562 \\
Electronic mail: rush@bonnie.astro.ucla.edu } 

\author { M. D. Bicay}
\affil{ S.I.R.T.F. Science Center, \\
Infrared Processing and Analysis Center, \\
Pasadena, CA, U.S.A. \\
Electronic mail: michael.d.bicay@jpl.nasa.gov } 

\author { B. O. Nelson}
\affil{ Infrared Processing and Analysis Center, \\
Pasadena, CA, U.S.A. \\
Electronic mail: nelson@ipac.caltech.edu } 

\author{R. M. Stanga}
\affil{ Universit\`a di Firenze \\
Istituto di Astronomia, Largo E. Fermi 5, I-50125 Firenze, Italy \\
Electronic mail:  stanga@arcetri.astro.it}

\and

\author { W. Webb}
\affil{ University of California \\
Department of Astronomy, 405 Hilgard Ave., \\
Los Angeles, CA, U.S.A. 90095-1562 \\
Electronic mail: webb@bonnie.astro.ucla.edu}

\begin{abstract}
We present 263 optical and near-infrared (NIR) images for 42 Seyfert 1s and 48
Seyfert 2s, selected from the Extended 12\,$\mu$m Galaxy Sample.
Elliptically-averaged profiles are derived from the images,
and isophotal radii and magnitudes are calculated from these.
We also report virtual aperture photometry, that
judging from comparison with previous work,
is accurate to roughly
0.05\,mag in the optical, and 0.07\,mag in the NIR.
Our $B$-band isophotal magnitude and radii, obtained from ellipse
fitting, are in good agreement with those of RC3.
When compared with the $B$ band, $V$, $I$, $J$, and $K$ isophotal diameters
show that the colors in the outer regions of Seyferts are consistent with 
the colors of normal spirals.
Differences in the integrated isophotal colors and comparison
with a simple model show that
the active nucleus$+$ bulge are stronger and redder
in the NIR than in the optical.
Finally, roughly estimated Seyfert disk surface brightnesses are
significantly brighter in $B$ and $K$ than those in normal spirals of 
similar morphological type.
\end{abstract}


\bigskip \bigskip \bigskip

\quad{\large
Accepted for publication in the Astrophysical Journal Supplement Series, 1999.
}

\section{Introduction}

Many studies have been devoted to the broadband optical and near-infrared
(NIR) properties of Seyfert nuclei and the galaxies that contain them.
Early photographic work was aimed at the large-scale morphology of
the host galaxies, and found
that, together with a surplus of disturbed and interacting systems, 
there may be a preponderance of annular structures
(Adams \cite{adams}). 
Unusually high outer ring fractions in Seyferts were also noted
by Simkin, Su, \& Schwarz (\cite{sss}), who suggested that the
patterns in Seyfert disk/ring morphology they observed could be 
explained by the consequences of radial gas inflow.
Peculiar or disturbed morphologies were also found in roughly a
third of the Seyfert sample studied by MacKenty (\cite{mackenty}).

Other early work noted a strong preference for peculiar
blue nuclei to reside in barred or weakly-barred spirals
(Sersic \& Pastoriza \cite{sp}; Sersic \cite{sersic}),
but this result was disputed by Heckman (\cite{heckman}) who
found that such nuclei occurred with equal frequency in barred 
and unbarred systems.
This was only the beginning of the search for bars in Seyferts, 
thought to be responsible via large-scale
gravitational torques for fueling active
nuclei (e.g., Shlosman, Begelman, \& Frank \cite{sbf}).
The latest studies show that while starbursts are almost always
barred (Hunt \& Malkan \cite{hm}), Seyferts are barred with
the same frequency as normal spirals
(McLeod \& Rieke \cite{mcleod}; Ho, Filippenko, \& Sargent (\cite{ho});
Mulchaey \& Regan \cite{mulchaey}).
Recently, though, it has been claimed that
bars and non-axisymmetric distortions occur
more frequently in Seyfert 2s than 1s (Maiolino et al. \cite{maiolino:1997}). 
Small-scale gaseous bars, or the ``bars-within-bars'' scenario proposed
by Shlosman, Frank, \& Begelman (\cite{sfb}) may be more relevant 
to the problem of nuclear fueling, because large-scale galactic bars 
cannot drive gas inwards on the small spatial scales 
of the galactic nucleus.
Indeed, NIR Hubble Space Telescope observations reveal 
a dusty nuclear bar about 80~pc long in a nearby Seyfert 
(Maiolino et al. \cite{maiolino}).

Later work was aimed at quantifying the broadband colors of 
the underlying galaxy and the Seyfert nucleus itself.
Although colors of host galaxies have been extensively 
studied, they are still a subject of debate.
Yee (\cite{yee}) and MacKenty (\cite{mackenty}) found optical
colors of Seyfert hosts to be similar to normal spirals,
while bluer integrated colors were found by Granato et al. (\cite{granato}),
and redder circumnuclear colors by Kotilainen \& Ward (\cite{kw}).
NIR colors are equally controversial:
Danese et al. (\cite{danese}), Kotilainen \& Ward (\cite{kw}),
and Alonso-Herrero, Ward, \& Kotilainen (\cite{herrero}) found
normal $J-H$ colors, but red $H-K$, and attributed these to 
hot dust and enhanced circumnuclear star formation. 
Normal $H-K$ colors, instead, were found by
Hunt \& Giovanardi (\cite{hg}) and Hunt et al. (\cite{hunt:1997}),
who corrected the observed colors for redshift.
Hunt et al. argued that the omission of this
correction could explain the red $H-K$ colors found by earlier work. 

Photometric studies of the Seyfert nucleus 
are hampered by the difficulty in separating the unresolved nuclear 
component from the underlying galaxy.
Nuclear amplitudes can be obtained by fitting a point-source
to the radial profiles (Yee \cite{yee}; McLeod \& Rieke \cite{mcleod}),
or by more sophisticated
decomposition techniques which fit one-dimensional 
brightness profiles with bulge, disk, and nuclear contributions
(Kotilainen et al. \cite{k92a}; Zitelli et al. \cite{zitelli}). 
It was suggested that the nuclei in type 1 Seyferts
are much bluer (optically) than type 2 nuclei, and that the ratio of 
nuclear-to-galaxy luminosity is much larger in type 1s (Yee \cite{yee}).
This last appears to be true also in the NIR, since
the starlight contribution in a given observing aperture was 
shown to be much larger in Seyfert 2s than in type 1s
(Alonso-Herrero et al. \cite{herrero}).

In this paper, we present new optical and NIR images of
Seyferts that can help understand and perhaps resolve some of the 
questions and controversies outlined above
(see also Peletier et al. \cite{peletier}).
These data are part of a larger study,
aimed at investigating structural features
in order to assess whether Seyfert galaxies are,
like their energetic nuclei, peculiar. 
The first paper in this series (Hunt \& Malkan \cite{hm}; hereafter Paper I) 
consisted of a statistical investigation of the morphological
properties of the 891 galaxies in the 12\,$\mu$m Galaxy Sample
(Rush, Malkan, \& Spinoglio \cite{rms}), including normal spirals,
starburst/HII galaxies, LINERs, and Seyferts.
We found that HII/starburst galaxies have a very high bar fraction,
but that Seyferts and LINERs have
the same frequency of bars as normal spirals.
Unlike starbursts, however, LINERs and Seyferts show rings significantly
more often than normal galaxies or starbursts:
LINERs have elevated rates of inner rings, and Seyferts have outer ring
fractions several times those in normal spirals.
Many of these results corroborate the suggestions of earlier work
(e.g., Adams \cite{adams}; Simkin et al. \cite{sss}),
but clearly a quantitative analysis of digital images is needed.

Here we present an optical and NIR image atlas of 90 
12\,$\mu$m-selected Seyferts, together with results from elliptical isophote fitting.
This set of 263 optical and NIR images for 42 Seyfert 1s and 48 Seyfert 2s
comprises 80\% of the 12\,$\mu$m Seyfert Sample, and is one of the largest image 
databases of Seyfert galaxies ever compiled.
Selection criteria of the Extended 12\,$\mu$m Galaxy Sample (E12GS) 
are described in Rush, Malkan, \& Spinoglio (\cite{rms}).
The following section evaluates how well the morphological properties 
of the observed sample represent the complete 12\,$\mu$m Seyfert Sample. 
Section \ref{obs} 
describes the observations, data reduction, and
photometric calibrations, as well as the aperture photometry.
Elliptical isophote fitting is presented in $\S$\,\ref{ell},
and the properties of the isophotal radii and magnitudes in {\it BVIJHK} are analyzed 
in $\S$\,\ref{iso}.
Section 6 discusses the Seyfert 
disk surface brightnesses, estimated from an exponential fit of the outer profiles.
An analysis of the properties of bars and lenses, 
in terms of their isophotal characteristics, 
among the different bands and between the
two Seyfert types is deferred to a companion paper.

\section{Properties of the Observed Sample\label{prop}\protect}

We are interested in the conformity of the 
observed subsample to the morphological properties of the
12\,$\mu$m Seyferts as a whole.
The presentation of the data in the following section
is therefore anticipated here by a preview of the
morphological characteristics of the observed subsample of 90 galaxies.
It turns out that
median morphological types, and bar and ring fractions of the 12\,$\mu$m 
Seyferts in this paper are very similar to the complete sample studied
in Paper I.
Seyfert 1s have a median type of T\,=\,1 (Sa), and 
Seyfert 2s T\,=\,2 (Sab).
65\% of the Seyfert 1s are strongly (SB) or weakly (SAB) barred,
as are 56\% of the Seyfert 2s.
40\% of the Seyfert 1s and 35\% of the Seyfert 2s have outer or pseudo-outer rings,
while 39\% (2s) to 45\% (1s) have inner rings.
Even the number ratio of type 2s to type 1s (48/42) is similar to that in
the complete sample (63/53).
Therefore, we conclude that the observed subsample provides a good representation
of the 12\,$\mu$m Seyferts as a whole, at least in terms of the morphology
we are studying.

The median recession velocities for the two types are comparable, but
with Seyfert 1s being generally farther away (median $z\,=\,0.021$)
than type 2s (median $z\,=\,0.015$). 
This difference, although moderate, could cause spurious differences to emerge
in type comparisons, and will be taken into account in the subsequent analyses.

\section{Observations, Reduction, and Photometry\label{obs}\protect}

Images of the sample galaxies were acquired over the course of several
observing campaigns at six different observatories: 
Calar Alto\footnote{The German-Spanish Astronomical Centre, Calar Alto, is
operated by the Max-Planck-Institute for Astronomy, Heidelberg, jointly
with the Spanish National Commission for Astronomy.}, 
CTIO\footnote{The Cerro Tololo Interamerican Observatory is operated by
the Association of Universities for Research in Astronomy, Inc., 
under a cooperative agreement with the National Science Foundation 
as part of the National Optical Astronomy Observatories.},
ESO\footnote{The European Southern Observatory is operated by a consortium
of eight European countries, with headquarters in Garching, Germany. }, 
Lick\footnote{Lick Observatory is operated by the University of
California Observatories.}, 
NOT\footnote{The Nordic Optical Telescope is operated on the island of 
La Palma jointly by Denmark, Finland, Norway, and Sweden, in the
Spanish Observatorio del Roque de los Muchachos of the Instituto de
Astrofisica de Canarias.}, and 
TIRGO\footnote{The Infrared Telescope at Gornergrat (Switzerland) is 
operated by CAISMI-CNR, Arcetri, Firenze.}.
Observations were performed with seven telescopes, 
five optical cameras, and three infrared ones.
Details of the observing runs and camera characteristics are given in
Table \ref{tbl:log}.
All filters are standard broad-band, and, when necessary, converted 
during calibration to well-defined photometric systems. 
Image reduction was performed in the IRAF 
environment\footnote{IRAF is the Image Analysis and Reduction Facility
made available to the astronomical community by the National Optical
Astronomy Observatories, which are operated by AURA, Inc., under
contract with the U.S. National Science Foundation.}, together 
with the STSDAS package\footnote{STSDAS is distributed by the Space
Telescope Science Institute, which is operated by the Association of
Universities for Research in Astronomy (AURA), Inc., under NASA contract
NAS 5--26555.}.

\begin{deluxetable}{lllclcl}
\tablecolumns{7}
\tableheadfrac{0.1}
\tablewidth{0pt}
\tablenum{1}
\tablecaption{Observing Runs and Camera Characteristics \label{tbl:log}}
\tablehead{
\colhead{Observatory} & \colhead{Telescope} & \colhead{Year} & 
\colhead{Camera} & \colhead{Pixels\tablenotemark{a}} & \colhead{Filters} & \colhead{Code} }
\startdata
Lick        & 1-m & 1990 & TI500  & 0.54 & {\it BI} & L90 \nl 
Lick        & 1-m & 1993 & Ford 2048$\times$2048 (binned)  & 0.37 & {\it V} & L93  \nl 
CTIO        & 0.91-m & 1992 & TK1024 (binned) & 0.792 & {\it BVI} & C92 \nl 
ESO         & Danish 1.54-m & 1992 & TK1024 & 0.38 & {\it BVI} & E \nl 
ESO         & MPI 2.2-m & 1992 & IRAC2 & 0.49 & {\it JHK} & I2 \nl 
TIRGO       & 1.5-m & 1992-93 & ARNICA  & 0.97 & {\it JHK} & T \nl 
Calar Alto  & 1.23-m & 1993 & TK510$\times$510 & 0.567 & {\it VI} & CA \nl 
Lick        & 1-m & 1993-94 & KCAM  & 0.50 & {\it K} & LK \nl 
NOT         & 2.56-m & 1995-96 & ARNICA  & 0.546 & {\it JHK} & N \nl 
\enddata
\tablenotetext{a}{In units of arcsec.} 
\end{deluxetable}

\subsection{Optical Data Reduction and Calibration}

Standard methods were used to reduce the optical images.
First, the bias was subtracted, using
when possible, the extended register overscan region. 
If not available, separate bias frames were acquired at the beginning and end of the night,
and the average of these was used for subtraction.
If necessary, 
dark frames at the same exposure time of the science frames were also subtracted. 
Pixel-to-pixel and large-scale spatial variations in the cameras$+$detectors
were corrected for with flat-field frames, acquired during twilight.
These were in most cases the combination of dome and sky flats, although
in some cases, only sky (ESO) or dome (Calar Alto) flats were used.
Cosmic rays were corrected for by the {\it cosmicrays} algorithm in the IRAF CCD 
reduction package, and bad columns with linear interpolation ({\it fixpix}).
The sky level was determined by averaging the median (over 5$\times$5 pixel boxes)
of several ($>$\,8--10) empty sky areas in the full-format CCD frame.

The reduced images were then calibrated with observations, obtained
throughout the various nights, of standard stars taken from the lists of Landolt
(\cite{landolt:1983}, \cite{landolt:1992}).
Each photometric measurement was the difference of a central aperture, typically
5 times the FWHM of the point-spread function (PSF), 
and a concentric, but larger, sky annulus.
For each filter,
a linear regression was performed on the run of zero points with airmass, and
when necessary to derive a color transformation, the residuals were fit to intrinsic color.
In some cases (Lick), such a procedure was not possible and mean extinction
and color terms were applied.
We estimate the formal photometric accuracy, as judged from the standard deviation
of the nightly standard stars, to be 0.02--0.03~mag or better for the Calar Alto,
CTIO, and ESO data; 0.05~mag for the Lick 1993 data; and
roughly 0.10--0.12~mag for the Lick 1990 run. 

\subsection{Near-Infrared Data Reduction and Calibration}

All three NIR cameras are based on 256$\times$256 HgCdTe arrays
(NICMOS3), and are described in more detail by
Lisi et al. (\cite{arnica:lisi}: ARNICA) and Hunt et al. (\cite{arnica:hunt}: ARNICA);
Moorwood et al. (\cite{irac2}: IRAC2); and Nelson et al (\cite{kcam}: KCAM).
In the ESO, TIRGO, and NOT campaigns, galaxies were observed by alternating
source and adjacent empty sky exposures, usually beginning and ending with
a sky frame, and integrating $\sim$\,1$^m$ in each position.
Typical total integration times were 5$^m$, although the NOT $K$-band observations
were usually twice that.
All three cameras employed a double-sampling read-out algorithm, so that the
bias level was automatically subtracted. 
The short on-chip integration times (10--30s) used to prevent saturation from 
the sky emission obviated the need for dark-current subtraction. 
Flat fields for each object exposure were obtained by averaging the sky
frames acquired immediately before and after it, after removal of any stars in
the sky frames.
For the TIRGO campaign, these were used directly. 
For the ESO (in $JHK$) and the NOT (in $K$ only) campaigns,
the average sky frame was first subtracted from the object exposure, 
then the difference divided by a differential flat field\footnote{The normalized
difference of a uniformly-illuminated frame and a dark one.}
obtained independently.
The (typically five) flat-fielded exposures were cleaned for bad pixels by 
interpolating a bad-pixel mask, then registered, and combined,
using a clipping procedure that relied on the (known) noise characteristics of the 
cameras.
Before combination, each exposure was rescaled to a common median level by adding
the appropriate constant.
The final step was the subtraction of the background that, as for the optical
images, was determined in empty sky regions by averaging the medians 
within several 5$\times$5 pixel boxes.

For the Lick KCAM campaign, the following procedure was used: 
each galaxy was exposed at seven different positions on the chip. 
A sky image was constructed for each position by taking the median
of the remaining six images. 
Each object image was then sky subtracted and corrected for the flat-field,
using dome flats taken at the end of each night. 
These frames were then registered and combined, as above.

The NIR images at ESO, NOT, and TIRGO were calibrated by observing stars 
from the UKIRT Faint Standard List (Casali \& Hawarden \cite{fs});
at Lick, they were calibrated with stars from Elias et al. (\cite{elias}).
As in the optical, each photometric measurement was
the difference of a central aperture, roughly 5 times the FWHM of the PSF,
and a concentric larger annulus which measured the sky.
For each run at ESO, NOT, and TIRGO, a fit for the atmospheric extinction coefficient 
and nightly zero points was performed, assuming that the extinction remained 
constant from night to night, but allowing the zero points (at unit airmass) to vary.
At Lick with KCAM, mean extinction coefficients were used, and only the zero points
were fit.
We found no evidence for a color transformation between the filter$+$HgCdTe camera 
combination and the InSb UKIRT Faint Star magnitudes 
(but see Hunt et al. \cite{arnica:std}).
Judging from the scatter of the standard star observations,
the formal photometric accuracy of the NIR data is roughly 
0.02~mag ($JHK$, ESO and NOT 1996), 0.03~mag ($K$, Lick),
0.05~mag ($JHK$, TIRGO), and
0.06--0.14~mag ($JHK$, NOT 1995).

\subsection{Quality Assessment}

\begin{figure}
\centerline{\epsfig{figure=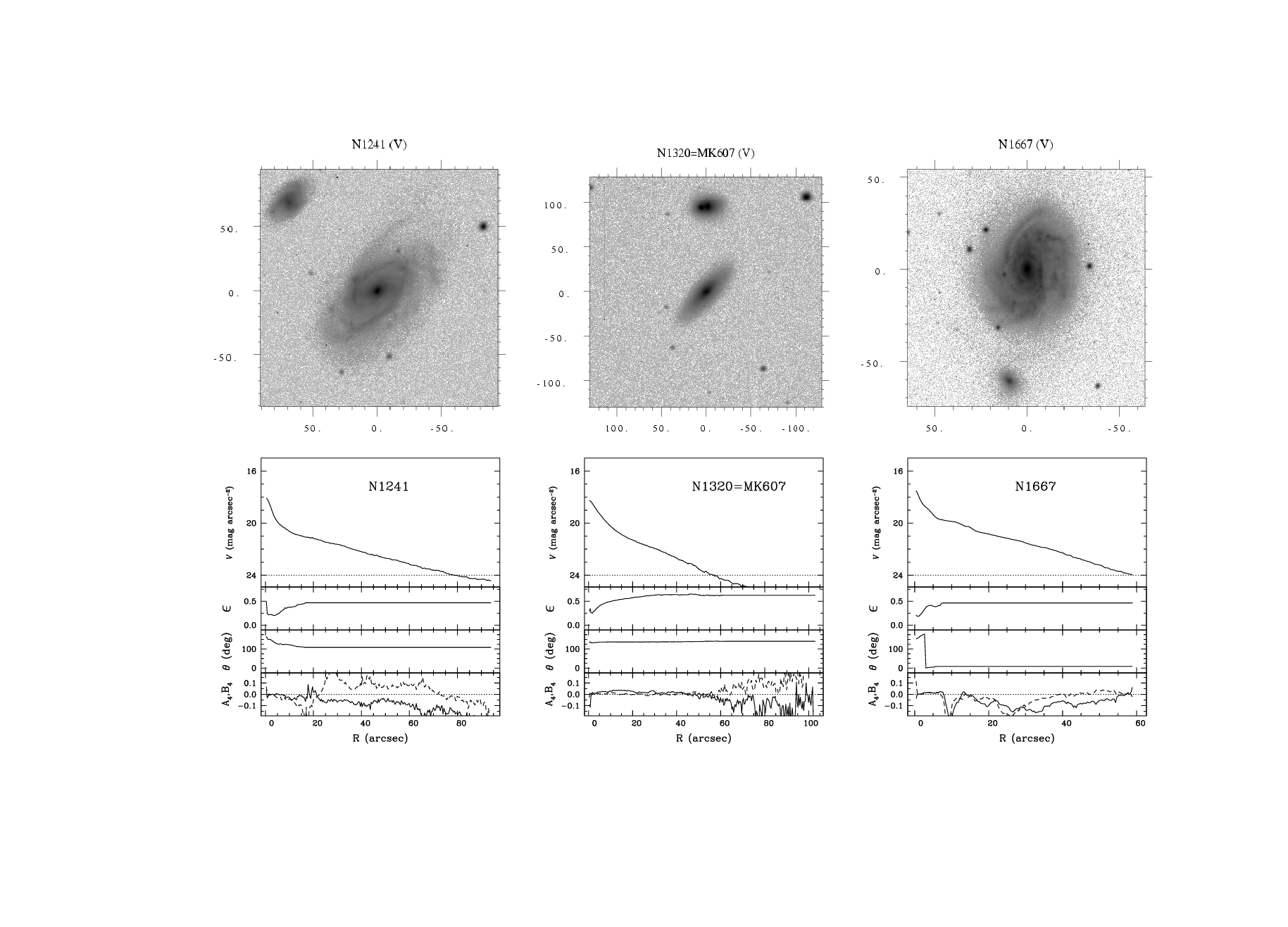,width=22cm,angle=90}}
\vspace{-2.0cm}
\figcaption[f1pub.ps]{Images and profiles (in $V$, $I$, $J$, or $K$)
of 90 12\,$\mu$m Seyferts, ordered in alphabetical order by name.
Images are in units of magnitudes/arcsec$^{-2}$, with grey scales
of 18.5 to 26.5 in $B$, 18 to 26 in $V$,
17 to 25.5 in $I$, 16 to 23.5 in $J$,
and 14.5 to 21.5 in $K$.
North is up, and East is the the left;
offsets are in arcsec from the nominal center, also used for profile
extraction.
The lower panel for each galaxy shows the radial brightness profile,
together with the radial runs of ellipticity, position angle, and
the $\sin\,4\theta$ coefficient $A_4$ (dashed line) 
and $\cos\,4\theta$ coefficient $B_4$ (solid line).
In the topmost of the lower panels, 
a dotted line shows fiducial isophotal levels, and in the lowermost, 
the nominal zero level. 
Shown here are representative examples only.
\label{fig:data}
} 
\end{figure}

Figure \ref{fig:data} shows for each of the 90 galaxies
an image and an elliptically-averaged profile (discussed in $\S$\,\ref{ell})
with radial runs of ellipticity, position angle, and the $\sin\,4\theta$
and $\cos\,4\theta$ coefficients, $A_4$ and $B_4$.
The virtual aperture photometry, 
isophotal magnitudes and radii (discussed in $\S$\,\ref{iso}),
and central surface brightnesses (discussed in $\S$\,\ref{exp}),
are given in
Table 2, 
whose entries are as follows: \\
\noindent {\it Column 1})~Seyfert type (integer). \\
\noindent {\it Column 2})~Source name or names. An asterix signifies
that the (radially varying) longer-wavelength best-fit ellipses were used to
derive the isophotal parameters.\\
\noindent {\it Column 3})~Morphological type (NED, RC3). \\
\noindent {\it Column 4})~Redshift (NED). \\
\noindent {\it Column 5})~Filter band. \\
\noindent {\it Column 6--8})~Aperture photometry with aperture diameters
of 10, 20, and 30\,arcsec, respectively. \\
\noindent {\it Column 9})~Magnitude integrated to the isophotal radius. \\
\noindent {\it Column 10})~Isophotal radius in arcsec,
evaluated at an isophote of 25\,mag\,arcsec$^{-2}$ in $B$;
24\,mag\,arcsec$^{-2}$ in $V$;
23\,mag\,arcsec$^{-2}$ in $I$;
22\,mag\,arcsec$^{-2}$ in $J$;
21.5\,mag\,arcsec$^{-2}$ in $H$; and 21\,mag\,arcsec$^{-2}$ in $K$. 
Radii marked with $^x$ are extrapolated to the isophote, instead of interpolated. \\
\noindent {\it Column 11})~Central surface brightness (mag\,arcsec$^{-2}$) 
obtained by fitting an exponential to the outer profile. \\
\noindent {\it Column 12})~Observing run code (see Col. 7 of Table \ref{tbl:log}). 
The code is "X" for two galaxies, NGC~2639 and NGC~3227, and signifies
images which were not acquired in the context of an observing run.
$J$ and $K$ images for NGC~2639 were taken at Palomar, and 
for NGC~3227 at Mt. Hopkins.\\

\begin{table}
\dummytable
\label{tbl:data}\protect
\end{table}

It can be seen from the Table 2 
that the wavelength coverage in the atlas is not uniform.
64 of the 90 galaxies observed have a $V$-band image, and this
is the band for which we have the most sources.
60 have a $K$-band image, and 35 (different sets of) objects
have $VK$ and $JK$. 
In each of the remaining optical and NIR bands, 
the wavelength coverage is similar, with
images for between 30 and 40 objects in each band $BIJH$.
Moreover, 
for each wavelength combination (e.g., $BV$, $BI$, $BVI$, $VK$, $JK$, $JHK$), 
there are 30 or more objects. 
Consequently, the statistics
should be sufficient for a detailed multiwaveband analysis.

To judge the effective spatial resolution of the atlas,
we have measured the PSFs of stars in the full-format CCD images 
(as opposed to the truncated versions shown in Fig. \ref{fig:data}).
The median full-width-half-maximum (FWHM) of the measured PSFs is 
2.0\,arcsec in $B$, 2.2 in $V$, 1.9 in $I$,
1.4 in $J$ and $H$, and 1.5\,arcsec in $K$.
Only 20 images of 263 have FWHM $>$ 3.0\,arcsec.
At the median redshifts of our sample (assuming $H_0\,=\,75$\,km/s/Mpc), 2\,arcsec
corresponds to a resolution of 600-800\,pc.
Given that typical bulge effective diameters in active galaxies are two--three
times this (Hunt et al. \cite{hunt:1999}), 
such resolution should be more than adequate
to separate the nuclear light from the bulge. 

The photometry in this paper can be checked by comparison
with previous results.
In the optical, we have relied on the variability study by
Winkler et al. (\cite{winkler}), conducted with a photometer,
with which we have 8 galaxies in common,
and on the imaging study by Kotilainen, Ward, \& Williger (\cite{kww})
with which we share 13.
If we omit from the comparison IC\,4329A which is known to vary 
in the optical (Winge et al. \cite{winge}),
our photometry is in good agreement with both groups, with a mean difference 
({\it us - them} over 8-10 data points)
of $-0.04\,\pm\,0.02$~mag in $B$,
$0.04\,\pm\,0.07$ in $V$, and
$0.075\,\pm\,0.04$ in $I$ relative to Winkler et al.,
and 
({\it us - them} over 12-20 data points)
$-0.04\,\pm\,0.06$~mag in $B$,
$-0.02\,\pm\,0.15$ in $V$, and
$0.07\,\pm\,0.06$ in $I$ relative to Kotilainen et al.
Three galaxies (NGC~4151, NGC~4593, 3C120)
contribute more than 0.2~mag to the latter comparison in $V$, one
of which is known to vary in the NIR (NGC~4151, McAlary et al. \cite{mcalary}).
Without these, we have $-0.002\,\pm\,0.04$ over 14 data points.
We also have compared our photometry for NGC\,1068 with the CCD study of
Schild, Tresch-Fienberg, \& Huchra (\cite{schild}), and find excellent
agreement (over the 10, 20, and 30\,arcsec apertures reported here)
with a mean difference ({\it us - them}) of $-0.02\,\pm\,0.006$~mag.

Comparison with other NIR studies is more problematic than in the optical
for several reasons outlined below.
Nevertheless, several of our objects have been observed by other authors
and, for the studies which have the most overlap
(10 each with Balzano \& Weedman \cite{bw} and Kotilainen et al. \cite{k92a}; 
14 galaxies with McAlary et al. \cite{mcalary}; 
36 with Spinoglio et al. \cite{spinoglio}), we have compared our
photometry with theirs.
In no case do we find significant systematic deviations; 
in all cases the rms difference is 0.15--0.25~mag.
One possible reason for such a large scatter may be
differences in the filter$+$detectors (Bessell \& Brett \cite{bessell}), 
a problem especially for the red nuclei in Seyferts.
Indeed, the observed rms scatter is reduced by roughly 40\% when
only $J-K\,<\,1.9$ are considered.
Another contributing factor may be
the limited chopper throw (typically a few to several tens of arcsecs) 
used to effect the sky subtraction in single-element photometry, since 
the mean difference {\it us\,--\,them} tends to be negative.
Small apertures and consequent difficulties with centering and seeing
constitute another problem; the scatter in all bands is reduced by a 
factor of two when only apertures $>$\,15\,arcsec are considered.

Intrinsic variability is almost certainly another factor since
many Seyfert galaxies vary in the NIR over timescales of years
(e.g., McAlary et al. \cite{mcalary};
Glass \cite{glass:1992}; Salvati et al. \cite{salvati}; Nelson \cite{nelson};
Glass \cite{glass:1997}; Glass \cite{glass:1998}), and
roughly half of the comparison objects are known or suspected variables
(McAlary et al. \cite{mcalary}).
Indeed scatter for
Seyfert 2s (thought to be less variable than Seyfert 1s, although see
Chelli et al. \cite{chelli}) is between a factor of 1.4 and 2
(in $K$ and $J$ respectively) lower than for both Seyfert types together.
Finally, the scatter among the papers we used for comparison
(and within them when measurements are conducted over several years) is similar
to the scatter between our values and theirs. 
All things considered, we
conclude that 0.1~mag is a conservative upper limit to
our NIR photometric errors, and that the NIR photometry is accurate 
to roughly 0.07~mag, similar to previous work.

\section{Elliptical Isophote Fitting \label{ell}\protect}

For each galaxy, in each band,
isophotes have been fit with ellipses using the STSDAS routine {\it ellipse}
in IRAF.
The center coordinates of each galaxy were determined by fitting a
two-dimensional Gaussian to the center [shown as (0,0) in Fig. \ref{fig:data}].
Then, keeping the center position fixed,
brightness profiles were derived in linear radial steps by letting the
surface brightness $\mu$, ellipticity $\epsilon\,=\,1-b/a$, and position
angle $\theta$ vary with radius.
Figure \ref{fig:data} illustrates these profiles (one band per galaxy),
as well as the $\sin\,4\theta$ and $\cos\,4\theta$ residuals of the best-fit ellipse. 
We have followed the IRAF convention in {\it ellipse} where $A_4$ refers
to the $\sin\,4\theta$ coefficient, and $B_4$ to $\cos\,4\theta$.

Of 64 galaxies with $B$ or $V$ {\it and}
$I$ or $J$ images, 16 appear to suffer from heavy extinction,
and the ellipse fits for the shorter wavelengths did not extend over 
a reasonable radius.
In these objects, 
we have fixed the radial runs of ellipticity and position angle of
the shorter wavelength profiles ($B$ or $V$) to the longer wavelength
ones and generated new profiles.
These sources are marked in Table 2 
with an asterix.

\begin{figure}
\centerline{\epsfig{figure=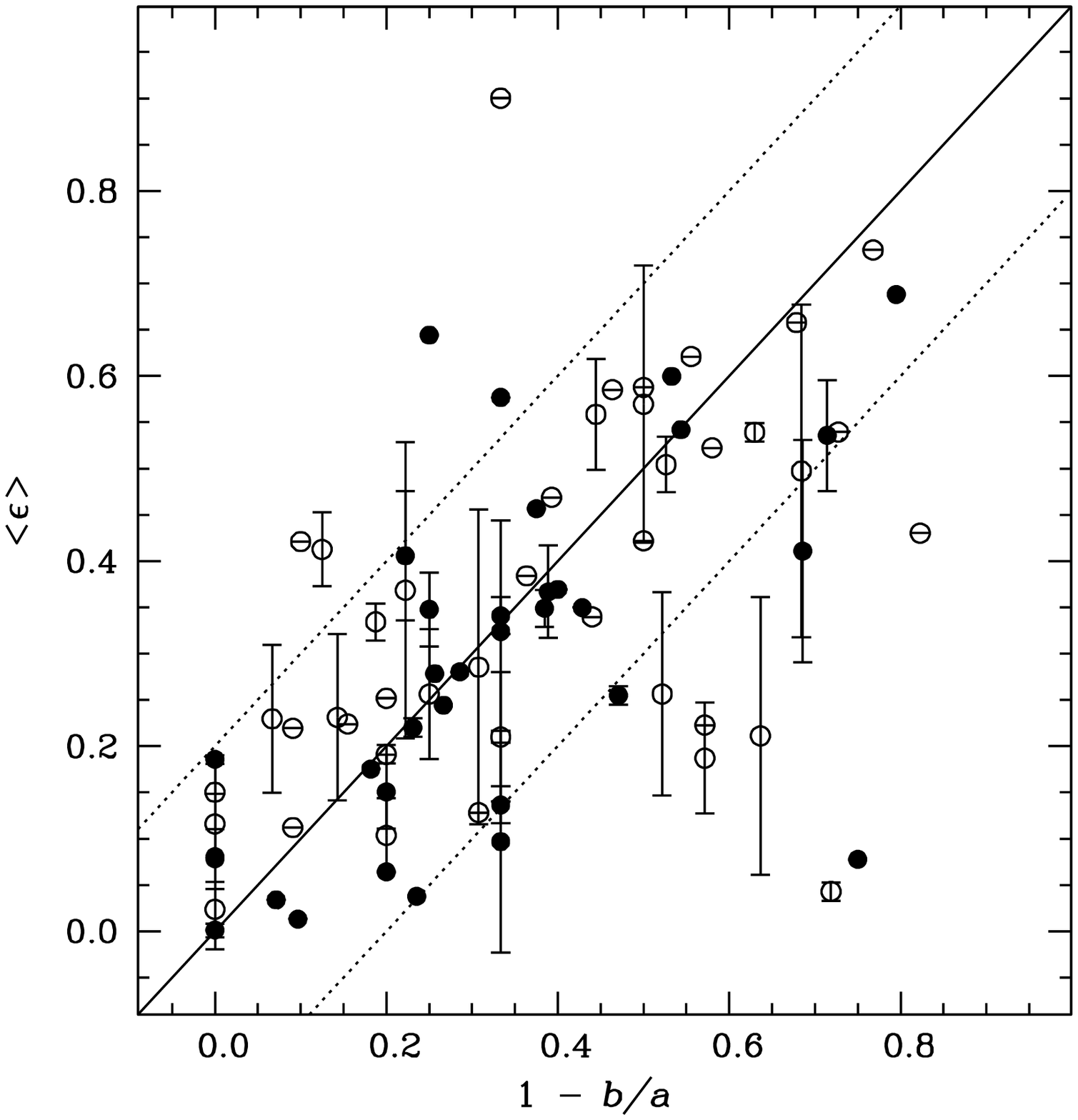,width=9cm}}
\figcaption[ellip.ps]{The mean ellipticity measured over $B$, $V$, $I$, and $J$
($K$ only when only band available) bands versus $1-b/a$ given in RC3.
Seyfert 1s are shown by filled circles, and Seyfert 2s by open ones.
Error bars are the standard deviation over the bands used in the mean calculation.
The solid line shows $< \epsilon >\,=\,1-b/a$, 
and the dotted lines enclose the region $|< \epsilon > - (1-b/a)| > 0.2$.
\label{fig:ell}
} 

\vspace{-3.0cm}
\centerline{\epsfig{figure=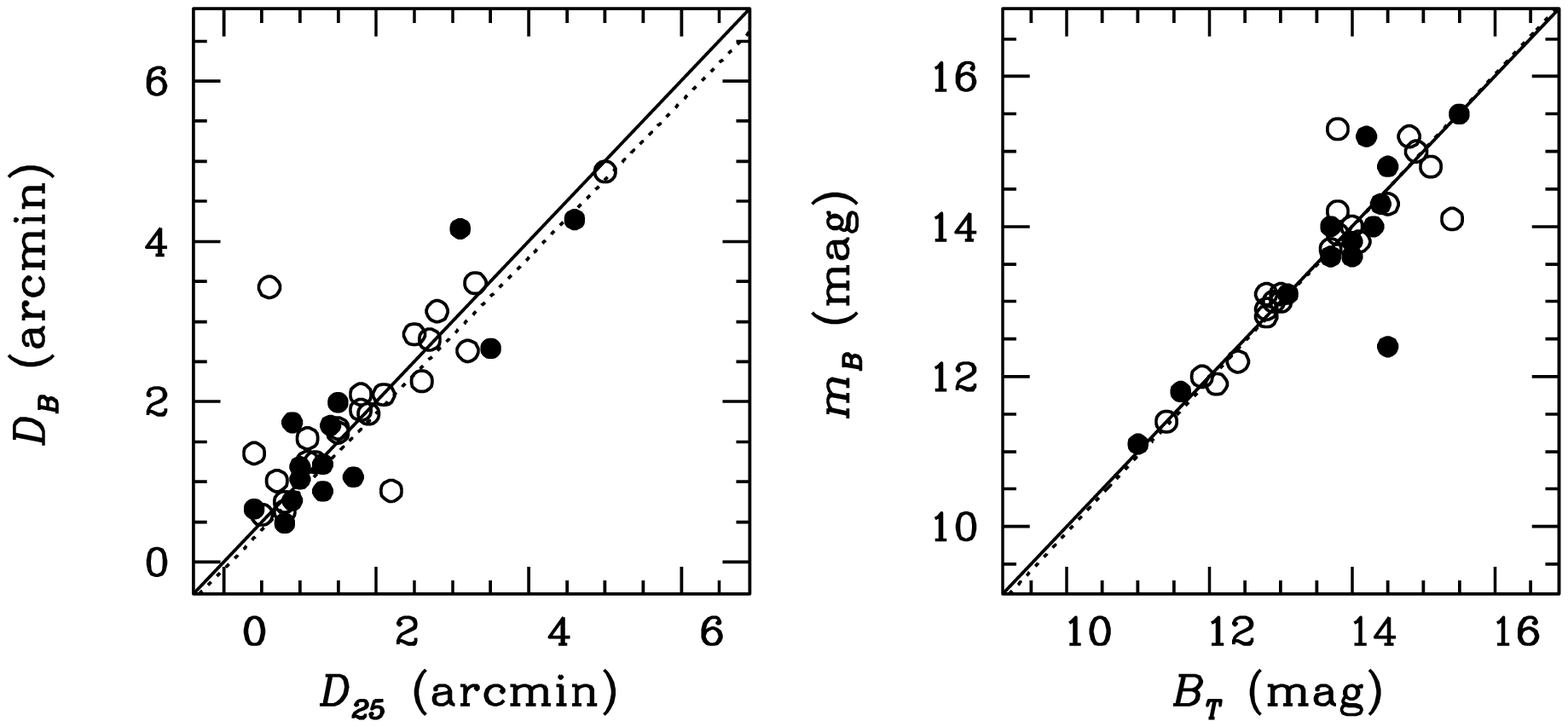,width=16cm}}
\vspace{-4.7cm}
\figcaption[radmag.ps]{Comparison of our $B$-band isophotal parameters
with those from RC3:
diameters are shown in the left panel, and magnitudes in the right.
Seyfert 1s are shown by filled circles, and Seyfert 2s by open ones.
Solid lines in both panels illustrate equality, and the dotted lines
show the best-fit regression.
The most conspicuous outlier in the left panel is Mrk~463, a 
double-nuclei galaxy with extrapolated isophotal parameters.
The two most conspicuous outliers in the right panel (below and
above the best-fit line, respectively) are Mrk~463 and Mrk~618, 
both with extrapolated radii.
\label{fig:radmag}
}
\end{figure}

Although an objective procedure, the ellipse fitting may be unduly influenced
by bars and spiral structure, and also, in the optical bands, by dust.
In those cases, we might expect values of ellipticity and position angle
to differ from those reported in catalogs.
A comparison of the ellipticities of the outer regions obtained from the 
ellipse fitting is shown in Fig. \ref{fig:ell}, where they are plotted against the 
equivalent values derived from the axial ratios given in 
the Third Reference Catalogue of Bright Galaxies (RC3: de Vaucouleurs et al. \cite{rc3}).
The ordinate axis in Fig. \ref{fig:ell} shows the mean $<\epsilon>$ over
the bands $BVIJ$, or $K$ only when only the $K$ band is available;
only the last twenty points in the profile are considered in the calculation of the mean.
The error bars correspond to a 1-$\sigma$ standard deviation over the
different bands available. 
It is evident from the figure that the scatter in the derived ellipticities 
is rather large; of 76 axial ratios available in RC3, 15 of our measured
ellipticities deviate from them
by more than 0.2 in either direction (indicated by the dotted lines).
Most discrepant ellipticities derived from the images are rounder than those in
RC3, implying that spiral structure and bars are not the most important cause
of the deviations.
Rather, it is likely that their cause is dust extinction, and its incapactitating
effect on the ellipse-fitting algorithm.
The isophotal values ($\S$\,\ref{iso}) for these galaxies could be
affected by the different ellipse parameters derived by the fitting algorithm,
but the availability of multiwavelength data will aid in the determination
of the correct values.

\section{Isophotal Diameters and Magnitudes in $BVIJHK$\label{iso}\protect}

Isophotal levels in the $BVIJHK$ bands were chosen to reflect the noise limits 
of the observations, but, more importantly, to conform to existing sources
of data.
Therefore, the $B$ fiducial isophote was defined to be 25\,mag\,arcsec$^{-2}$ in
accordance with RC3.
The $H$ limit of 21.5\,mag\,arcsec$^{-2}$ is the same as that used in two
extensive studies of normal spiral galaxies in the Coma
(Gavazzi et al. \cite{gavazzi}) and Pisceus-Perseus superclusters
(Moriondo et al. \cite{moriondo}).
The intervening isophotal limits were defined to follow a smooth
trend with $\log\,\lambda$:
$\mu_{iso}(V)\,=\,24$\,mag\,arcsec$^{-2}$,
$\mu_{iso}(I)\,=\,23$\,mag\,arcsec$^{-2}$,
$\mu_{iso}(J)\,=\,22$\,mag\,arcsec$^{-2}$, and \\
$\mu_{iso}(K)\,=\,21$\,mag\,arcsec$^{-2}$ in $K$.

The radius $R_{{\it band}}$ at a given isophote is the semi-major axis of the 
fitted elliptical isophote at the isophotal magnitude, and they are reported in 
Table 2. 
$R_{\it band}$ was determined usually by linear interpolation, 
that is by fitting an exponential to the outer isophotes.
In a minority of cases (32/263), mostly for the apparently large galaxies,
extrapolation was necessary to define $R$ since the profile did not
reach the level defined by the isophote; these are marked with $x$ in 
Table 2. 
Isophotal magnitudes $m_{\it band}$ were derived by integrating the radial
surface brightness profile from the center out to $R_{\it band}$, or to the last
measured point when $R_{\it band}$ had to be extrapolated.
These values are also given in Table 2. 

A comparison of our $B$-band isophotal parameters and those in
RC3 is shown in Fig. \ref{fig:radmag}.
The solid lines in both panels correspond to equality, and
the dotted lines show the best-fit regression\footnote{Having
eliminated the outlier Mrk~463 from the diameter relation, since it has
double nuclei, does not appear to be well-approximated by
the ellipse fitting, and is an extrapolated value.}
using the ordinary least-squares (OLS) bisector advocated by
Isobe et al. (\cite{isobe}).
Both diameters and magnitudes are in good agreement with those in RC3, and the fitted
regressions are compatible with unit slope.
The most discrepant point in the right panel is Mrk~618 which we find
more than 2~mag brighter than the RC3 value; 
its magnitude is derived from an extrapolated value of R$_{B}$, and
is more than twice as large as those in $VJHK$, 
which are consistent among themselves and with fainter magnitudes.
The next most discrepant point is  Mrk~463 (see footnote), 1.5~mag 
fainter than the RC3 value, and the magnitude is
also derived from an extrapolated R$_{B}$.

\subsection{Trends With Isophotal Diameters and Colors\label{trends}\protect}

\begin{figure}
\centerline{\epsfig{figure=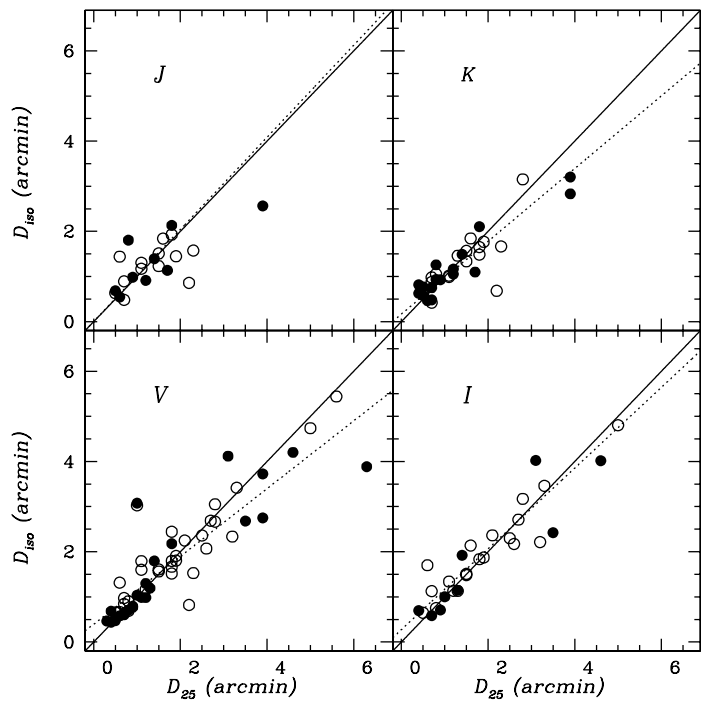,width=10cm}}
\figcaption[radcol.ps]{Comparison of our $VIJK$-band isophotal diameters
with those from RC3.
Seyfert 1s are shown by filled circles, and Seyfert 2s by open ones.
As in Fig. \ref{fig:radmag}, solid lines correspond to equality, and
the dotted lines to the best-fit regression.
The less-than-unit slopes in all but the $J$ band imply that the
outer colors are bluer than the arbitrary
isophotal fiducial magnitude differences listed in the text.
\label{fig:radcol}
}
\end{figure}

Figure \ref{fig:radcol} illustrates the comparison of our isophotal
diameters in $VIJK$ with those in $B$ from RC3.
As in Fig. \ref{fig:radmag}, the solid lines correspond to equality,
and the dotted lines show the best-fit regression.
In $V$, $I$, and $K$, the best-fit slopes are less than one,
(0.75 in $V$, 0.81 in $K$, and 0.89 in $I$), but in $J$ the fitted
slope (1.02) is approximately unity. 
The less-than-unit slopes in $V$, $I$, and $K$ are expected
when the corresponding
colors in the outer regions ($B-V$, $B-I$, $B-K$) 
are bluer than the difference in the fiducial isophotes
[$(\mu_B-\mu_V)_{iso}\,=\,1$, $(\mu_B-\mu_I)_{iso}\,=\,2$, $(\mu_B-\mu_K)_{iso}\,=\,4$].
This finding implies that colors of Seyfert disks are similar to those
of normal early-type spirals, since De Jong (\cite{dejong:iv}) 
reports integrated early-type spiral
colors of $B-V\,=\,0.78$, $B-I\,=\,1.81$, and $B-K\,=\,3.65$. 

\begin{figure}
\centerline{\epsfig{figure=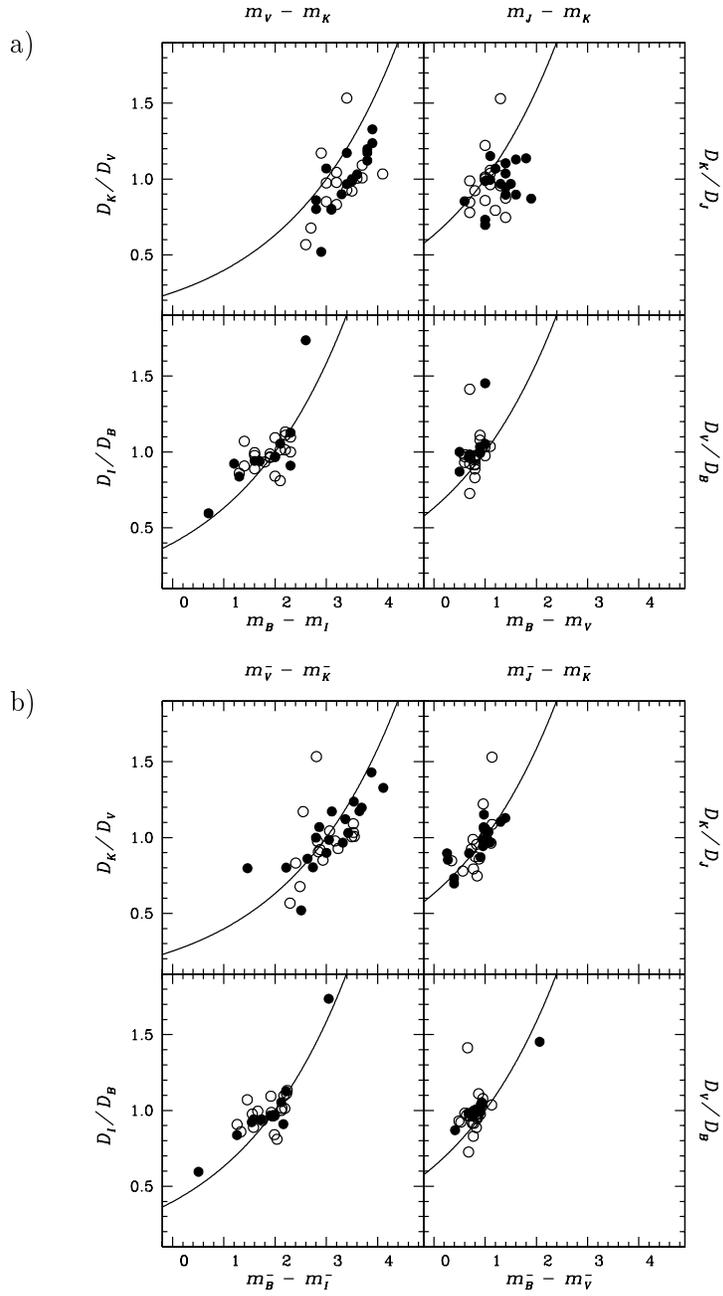,width=16cm}}
\vspace{-2.0cm}
\figcaption[figcol.ps]{Isophotal diameter ratios versus
integrated isophotal colors for a restricted set of color combinations.
Seyfert 1s are shown by filled circles, and Seyfert 2s by open ones.
The solid line illustrates the behavior of a pure exponential disk with
the total luminosity given by the isophotal magnitude, and
with a central color equal to the difference
in the isophotal fiducial magnitudes.
Panel a) shows the simple difference of the isophotal magnitudes;
panel b) shows the same difference, but after the flux from
the central 10\,arcsec has been subtracted.
\label{fig:col}
}
\vspace{-2.0cm}
\end{figure}

We can further investigate the colors of the outer regions
by comparing the ratios of diameters in two different bands to the
integrated isophotal color. 
Such comparisons are shown in  Fig. \ref{fig:col}a.
The solid curves in all panels represent an exponential disk
with total disk luminosity given by the isophotal magnitudes, and 
the central color 
(difference in disk central surface brightness $\mu_d$ in the two bands)
given by the difference in the fiducial isophotes.
When this last is true, the ratio of the isophotal diameters is 
equal to the disk scale-length ratio, and this ratio, in principle, 
depends only on the total color.
The previous paragraph demonstrated, however, that 
for $B-V$, $B-I$, and $B-K$, the outer colors
tend to be bluer than the difference in the fiducial isophotes.
In this case, the ratio of the isophotal diameters becomes a
complicated function of disk scale-length ratio and central color.
Either way, Fig. \ref{fig:col}a shows that $B-I$ and $B-V$ have a 
relatively small scatter, and are consistent with integrated colors of
normal early-type spirals.
An examination of the mean isophotal colors confirms this with
$<B-V>\,=\,0.71\,\pm\,0.45$, $<B-I>\,=\,1.85\,\pm\,0.42$. 
The optical--NIR colors, on the other hand, show a larger spread and tend to be 
redder than those expected for normal spirals
($<B-K>\,=\,4.23\,\pm\,0.85$), in contrast with
what we inferred in the previous paragraph.

Because the isophotal magnitudes we measure also contain the
active nucleus and the bulge, assuming that they reflect only an exponential
disk is incorrect.
Active nuclei and bulges are both ``red'' contributions to $V-K$ and $J-K$, and 
would be expected to pull the isophotal colors to the right in Fig. \ref{fig:col}.
For example, a normal disk should have $V-K$ colors of $\approx\,3.0$, but
almost all the isophotal colors shown in the upper left panel of Fig. \ref{fig:col}a 
are redder than this, although they follow the general trend shown by
the solid curve.
The data in both upper panels of Fig. \ref{fig:col}a also show a larger dispersion
than the lower panels. 
Both of these considerations suggest that, relative to the disk, 
the emission from the active nucleus$+$bulge is more significant 
{\it and} redder 
at wavelengths longer than 1\,$\mu$m than in the optical. 

We can verify the validity of such a claim by subtracting the central flux
from the isophotal magnitude, since the ``red'' contribution is confined to the
inner regions.
This has been done using the 10-arcsec aperture photometry 
reported in Table 2, 
and the resulting plots are shown in Fig. \ref{fig:col}b.
Although the 10-arcsec flux is only a crude estimate of the 
active nucleus$+$bulge, Fig. \ref{fig:col}b supports
the hypothesis that their contribution relative to the disk
is stronger and redder at longer wavelengths.
In the optical bands, the scatter is similar to the photometry before subtraction,
but in the NIR, the data after subtraction follow the exponential disk 
prediction very well, and the scatter relative to it is reduced significantly.

\section{Exponential Surface Brightnesses in $B$ and $K$\label{exp}\protect}

The surface brightness of Seyfert disks has been the subject
of some controversy.
By fitting an exponential to the radial profiles, 
Yee (\cite{yee}) tentatively concluded that Seyfert disks were
brighter than normal spirals, although noted that selection effects
and problems with nuclear subtraction could be responsible for the
difference.
With a modified version of the same method,
similar to what we have done here (see below),
MacKenty (\cite{mackenty}) found no such difference in disk surface 
brightness.
With two-dimensional NIR bulge$+$disk$+$nucleus decompositions
Hunt et al. (\cite{hunt:1999}) found that, for a given disk radius,
Seyfert disks were almost 1~mag/arcsec$^2$ brighter at 2\,$\mu$m than their 
normal counterparts,
more similar to bulges than to ordinary spiral disks.

Because of this disagreement regarding the properties of
Seyfert disks, we discuss briefly
the disk surface brightnesses obtained 
by fitting an exponential to the outer isophotes.
(This is the same fit used to determine the
isophotal parameters, as described in $\S$\,\ref{iso}.)
However, as pointed out by Mihalas \& Binney (\cite{mb}), the bulge
contribution is nonnegligible even at large radii,
making the slope of the derived disk component smaller than the
slope of the tangent to the galaxy profile.
Moreover, one-dimensional profile decomposition can be inaccurate
even when the bulge is included in the fit (Byun \& Freeman \cite{byun}).
In spite of these caveats, we have reported the
surface brightnesses derived in this way in Table 2 
(Col. 11); they have not been corrected for inclination, 
and should be considered rough estimates ($\pm\,0.5$~mag).

We have compared the Seyfert surface brightnesses with those of
normal spirals taken from de Jong (\cite{dejong:iii})\footnote{Also not 
corrected for inclination.}.
The mean surface brightnesses of the two samples\footnote{Over objects with 
well-defined morphological types.} are: 21.44 $B$- and 17.48 $K$-mag/arcsec$^2$ 
for normal spirals, and 20.51 $B$- and 16.80 $K$-mag/arcsec$^2$ for the Seyferts.
The normal value is averaged only over the spiral types present in the
Seyfert sample (T\,$\leq$\,5) (see below).
Figure \ref{fig:disk} shows the surface brightnesses in $B$ and
$K$ plotted against morphological type T;
Seyferts are shown as filled symbols (1s as circles, 2s as squares)
and normal spirals by $\times$s.
The sample means are shown by dotted (Seyferts) and dashed (normal) lines 
in the figure.
We have plotted the surface brightnesses as a function of T since there is 
some indication that, at least in $K$, disks increase in surface brightness going
from late- (Sbc-Sc) to early (Sa-Sab) types
(de Jong \cite{dejong:iii}; Hunt et al. \cite{hunt:1999};
Moriondo et al. 1999, in preparation).
This could be a critical point since 
Seyferts are usually early types, and, if compared to all spiral types or 
later ones, their disks could appear to be brighter simply because of their 
Hubble type.

In any case, Seyfert disks appear to be brighter in both $B$ and $K$
than those in normal spirals of similar morphological type.
A formal comparison of the means and spreads (shown in Fig. \ref{fig:disk})
gives a probability in
both bands of $>$ 99.9\% that the two samples are distinct.
This result is clearly not conclusive because of the caveats outlined
above, and will be the subject of a future paper based on
two-dimensional bulge$+$disk$+$nucleus decompositions.

\section{Summary\label{summary}\protect}

We have presented 263 optical and NIR images for 42 Seyfert 1s and 48
Seyfert 2s, selected from the Extended 12\,$\mu$m Galaxy Sample.
Judging from comparison with previous work,
the aperture photometry derived from the images is accurate to roughly
0.05\,mag in the optical, and 0.07\,mag in the NIR.
$B$-band isophotal magnitudes and radii, obtained from ellipse fitting,
are in good agreement with those of RC3.
When compared with the $B$ band, $V$, $I$, $J$, and $K$ isophotal diameters
show that the colors in the outer regions are consistent with the colors 
of normal spirals, in accordance with previous studies.
Differences in the integrated isophotal colors and comparison
with a simple model show that, relative to the disk,
the active nucleus$+$ bulge emission is larger and redder at longer wavelengths.
Finally, roughly estimated Seyfert disk surface brightnesses are
significantly brighter than those in normal spirals of similar morphological type.

Future work based on this image atlas includes: 
1)~an analysis of the radial runs of ellipticity,
position angle, and quadrupole terms for a quantitative determination
of bar and lens properties in Seyferts;
2)~two-dimensional decomposition into bulge, disk, and active nucleus;
and 3)~an analysis of the color images with particular emphasis on the
long-wavelength leverage color $V-K$, for which we have the most sources.

\acknowledgements
As ever, we would like to thank C. Giovanardi for insightful harassment.
One of us (MAM) would like to thank the Italian National Astronomy Group
(G.N.A. -- C.N.R.)
and the Osservatorio Astrofisico di Arcetri for financial support.
We are grateful to S. Sakai for performing the NIR observations at Palomar
and Mt. Hopkins, and
to A. Richichi for sharing his observing time at Calar Alto.
This research was partially funded by ASI Grant ARS-98-116/22, and made
use of the NASA/IPAC Extragalactic Database (NED), operated by the
Jet Propulsion Laboratory,
California Institute of Technology, under contract with the U.S.
National Aeronautics and Space Administration.

\newpage
\onecolumn

\begin{figure}
\centerline{\epsfig{figure=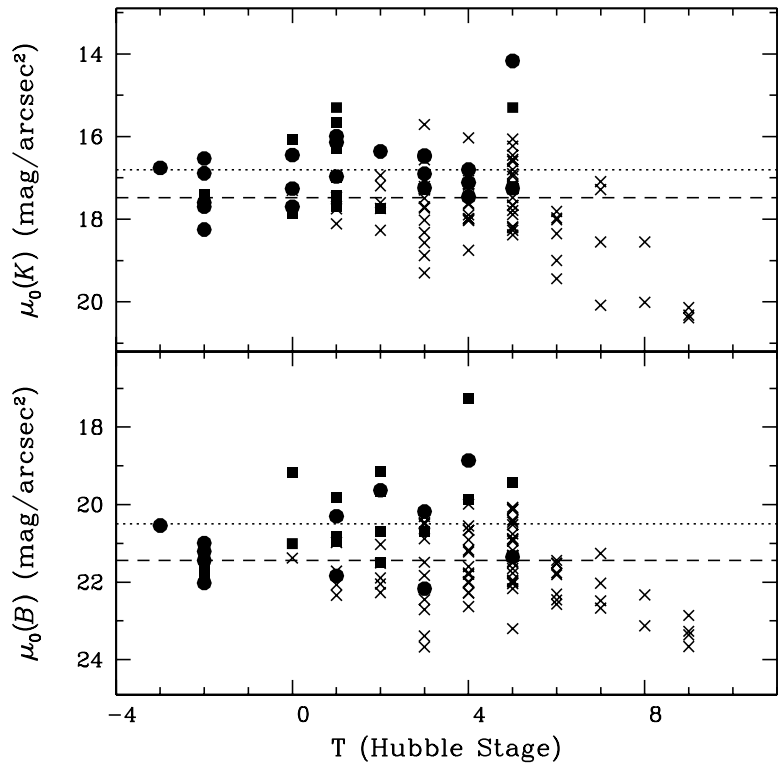,width=12cm}}
\figcaption[disk.ps]{The surface brightness $\mu$ derived by an
exponential fit to the outer profiles plotted against morphological
type T.
The upper panel shows the $K$ band, and the lower $B$.
Seyfert 1s are shown by filled circles, Seyfert 2s by filled squares,
and normal spirals (de Jong cite{dejong:iii}) by $\times$s.
Neither data set has been corrected for inclination.
The mean Seyfert surface brightness for the galaxies plotted
here (i.e., those with well-defined morphological type) is shown by the
dotted lines, and dashed lines show the mean for normal spirals with T\,$\leq$\,5.
\label{fig:disk}
}
\end{figure}

\end{document}


\begin{deluxetable}{clllrrrrrrll} 
\tablecolumns{12} 
\tableheadfrac{0.05} 
\small
\tablewidth{0pt}
\tablenum{2}
\tablecaption{12\,$\mu$m Seyfert Aperture and Isophotal Photometry \label{tbl:data}}
\tablehead{ 
\multicolumn{1}{c}{Type} & \multicolumn{1}{c}{Name } & \multicolumn{1}{c}{Morph.} &
\multicolumn{1}{c}{$z$} &  \multicolumn{1}{c}{Band} &
\multicolumn{1}{c}{10"} &  \multicolumn{1}{c}{20"} &  \multicolumn{1}{c}{30"} &
\multicolumn{1}{c}{$m_{\it band}$} &  \multicolumn{1}{c}{$R_{\it band}$} & \multicolumn{1}{c}{$\mu_{\it disk}$} & \multicolumn{1}{c}{Run} \\
\colhead{(1)} &  \colhead{(2)} & 
\colhead{(3)} & \colhead{(4)} &
\colhead{(5)} & \colhead{(6)} &
\colhead{(7)} & \colhead{(8)} &
\colhead{(9)} & \colhead{(10)} & \colhead{(11)} & \colhead{(12)} } 
\startdata
 1 & 3C120 & S0: & 0.0330 & $ B$  &  15.27 &  15.18 &  15.13 &  15.16 &   14.5 &   21.4 & E\nl
   &  &  &        & $ V$  &  14.55 &  14.37 &  14.26 &  14.21 &   21.1 &   21.0 & E\nl
   &  &  &        & $ K$  &  10.74 &  10.56 &  10.45 &  10.31 &   28.0 &   17.6 & I2\nl
 1 & 3C234 &  & 0.1848 & $ K$  &  12.71 &  12.56 &  12.45 &  12.44 &   22.5 &   20.5 & T\nl
 1 & 3C445 &  & 0.0562 & $ J$  &  13.31 &  13.21 &  13.15 &  13.23 &    9.0 &   18.9 & N\nl
   &  &  &        & $ H$  &  12.23 &  12.13 &  12.07 &  12.13 &   11.4 &   18.5 & N\nl
   &  &  &        & $ K$  &  11.31 &  11.27 &  11.25 &  11.28 &    7.9 &   17.0 & N\nl
 2 & ARP220=UGC09913 & S?+Sm & 0.0181 & $ J$  &  12.43 &  11.71 &  11.41 &  11.02 &   36.9 &   18.1 & N\nl
   &  &  &        & $ K$  &  10.84 &  10.34 &  10.11 &   9.74 &   40.1 &   17.3 & N\nl
 2 & CGCG381-051 &  & 0.0306 & $ J$  &  13.66 &  12.84 &  12.46 &  12.34 &   22.0 &   17.3 & N\nl
   &  &  &        & $ H$  &  12.95 &  12.15 &  11.77 &  11.64 &   25.6 &   17.3 & N\nl
   &  &  &        & $ K$  &  12.68 &  11.96 &  11.62 &  11.53 &   20.3 &   16.3 & N\nl
 1 & ESO141-G055 & S? & 0.0360 & $ B$  &  14.41 &  14.26 &  14.19 &  14.02 &   23.1 &   21.6 & C92\nl
   &  &  &        & $ V$  &  14.04 &  13.82 &  13.74 &  13.51 &   23.1 &   21.1 & C92\nl
   &  &  &        & $ I$  &  13.40 &  13.10 &  12.99 &  12.82 &   21.3 &   19.6 & C92\nl
 2 & ESO541-IG012 &  & 0.0566 & $ J$  &  13.35 &  13.01 &  12.62 &  12.14 &   28.7 &   19.4 & N\nl
   &  &  &        & $ H$  &  12.21 &  11.98 &  11.68 &  11.29 &   26.8 &   18.6 & N\nl
   &  &  &        & $ K$  &  11.25 &  11.12 &  10.95 &  10.71 &   21.4 &   18.1 & N\nl
 2 & F01475-0740 & E-S0 & 0.0177 & $ V$  &  15.97 &  15.71 &  15.64 &  15.73 &   10.1 &   20.1 & L93\nl
   &  &  &        & $ J$  &  13.74 &  13.56 &  13.48 &  13.59 &   10.1 &   18.4 & N\nl
   &  &  &        & $ H$  &  12.94 &  12.77 &  12.71 &  12.77 &   10.9 &   18.0 & N\nl
   &  &  &        & $ K$  &  12.44 &  12.30 &  12.24 &  12.31 &   15.5 &   19.2 & N\nl
 2 & F03362-1642 &  & 0.0369 & $ B$  &  17.67 &  17.04 &  14.18 &  14.14 &   19.1 &   22.1 & E\nl
   &  &  &        & $ V$  &  16.39 &  15.80 &  13.49 &  13.45 &   27.0 &   20.9 & E\nl
   &  &  &        & $ K$  &  11.62 &  11.38 &  10.74 &  10.50 &   31.7 &   18.4 & I2\nl
 1 & F03450+0055 &  & 0.0310 & $ V$  &  14.69 &  14.59 &  14.56 &  14.59 &    9.3 &   19.3 & L93\nl
   &  &  &        & $ K$  &  10.75 &  10.68 &  10.66 &  10.68 &   11.6 &   16.6 & LK\nl
 2 & F04385-0828 & S0 & 0.0151 & $ V$  &  15.70 &  15.14 &  14.95 &  14.93 &   19.9 &   20.0 & L93\nl
 2 & F05189-2524 & pec & 0.0426 & $ V$  &  16.06 &  15.72 &  15.60 &  15.72 &   11.0 &   20.9 & E\nl
   &  &  &        & $ I$  &  16.22 &  15.91 &  15.79 &  16.10 &    6.8 &   20.3 & E\nl
 1 & F05563-3820 &  & 0.0344 & $ V$  &  15.18 &  14.97 &  14.90 &  14.93 &   18.1 &   20.4 & E\nl
   &  &  &        & $ I$  &  13.80 &  13.60 &  13.50 &  13.52 &   17.5 &   18.6 & E\nl
   &  &  &        & $ K$  &  11.98 &  11.82 &  11.76 &  11.81 &   14.5 &   18.1 & I2\nl
 1 & F07599+6508 &  & 0.1488 & $ K$  &  10.38 &  10.26 &  10.22 &  10.20 &   25.3 &   18.8 & LK\nl
 2 & F08572+3915 &  & 0.0582 & $ V$  &  17.04 &  16.42 &  16.24 &  16.44 &   11.0 &   20.7 & L93\nl
 1 & F13349+2438 &  & 0.1076 & $ K$  &  10.60 &  10.44 &  10.41 &  10.42 &   13.3 &   15.5 & LK\nl
 1 & F15091-2107$^*$ &  & 0.0446 & $ B$  &  16.00 &  15.74 &  15.52 &  15.50 &   19.8 &   22.3 & C92\nl
   &  &  &        & $ V$  &  15.10 &  14.81 &  14.61 &  14.59 &   20.5 &   21.4 & C92\nl
   &  &  &        & $ I$  &  13.96 &  13.66 &  13.48 &  13.43 &   20.9 &   20.1 & C92\nl
   &  &  &        & $ K$  &  11.30 &  11.06 &  10.93 &  10.83 &   24.5 &   18.3 & LK\nl
 2 & F19254-7245 &  & 0.0617 & $ B$  &  17.39 &  16.60 &  16.27 &  16.28 &   17.2 &   21.5 & C92\nl
   &  &  &        & $ V$  &  16.37 &  15.61 &  15.29 &  15.32 &   17.8 &   20.5 & C92\nl
   &  &  &        & $ I$  &  14.92 &  14.26 &  13.98 &  13.99 &   18.9 &   19.1 & C92\nl
 2 & F22017+0319 &  & 0.0611 & $ J$  &  13.48 &  13.17 &  13.03 &  13.06 &   14.4 &   18.8 & N\nl
   &  &  &        & $ H$  &  12.77 &  12.48 &  12.35 &  12.39 &   14.1 &   17.9 & N\nl
   &  &  &        & $ K$  &  11.97 &  11.79 &  11.72 &  11.73 &   12.6 &   17.5 & N\nl
 1 & IC4329A & SA0+:sp & 0.0161 & $ B$  &  14.84 &  14.48 &  14.28 &  13.83 &   51.1 &   21.0 & C92\nl
   &  &  &        & $ V$  &  13.77 &  13.42 &  13.23 &  12.84 &   53.8 &   20.3 & C92\nl
   &  &  &        & $ I$  &  12.31 &  12.02 &  11.85 &  11.48 &   57.7 &   19.1 & C92\nl
 2 & IC5063 & SA(s)0+: & 0.0113 & $ B$  &  15.16 &  14.39 &  13.99 &  13.02 &   62.6 &   21.5 & C92\nl
   &  &  &        & $ V$  &  14.10 &  13.36 &  12.98 &  12.05 &   67.5 &   20.8 & C92\nl
   &  &  &        & $ I$  &  12.77 &  12.03 &  11.66 &  10.76 &   70.9 &   19.5 & C92\nl
\tablebreak
 1 & IZw001 & S? & 0.0611 & $ V$  &  14.25 &  14.12 &  14.03 &  14.04 &   14.2 &   20.3 & L93\nl
   &  &  &        & $ J$  &  12.30 &  12.16 &  12.06 &  12.06 &   15.8 &   18.5 & N\nl
   &  &  &        & $ H$  &  11.35 &  11.23 &  11.15 &  11.14 &   17.1 &   17.9 & N\nl
   &  &  &        & $ K$  &  10.59 &  10.52 &  10.48 &  10.49 &   14.2 &   17.1 & N\nl
 2 & MCG+00-29-023 & SAB(s)b & 0.0249 & $ B$  &  16.20 &  15.62 &  15.38 &  15.25 &   22.6 &   20.7 & C92\nl
   &  &  &        & $ V$  &  15.17 &  14.66 &  14.44 &  14.36 &   20.6 &   19.6 & C92\nl
   &  &  &        & $ I$  &  13.67 &  13.22 &  13.05 &  12.94 &   22.6 &   18.7 & C92\nl
 2 & MCG-02-40-004=NGC5995 &  & 0.0252 & $ B$  &  15.39 &  14.86 &  14.59 &  14.35 &   30.4 &   20.9 & C92\nl
   &  &  &        & $ V$  &  14.44 &  13.92 &  13.68 &  13.46 &   29.2 &   20.0 & C92\nl
   &  &  &        & $ I$  &  13.05 &  12.56 &  12.33 &  12.08 &   33.8 &   19.3 & C92\nl
   &  &  &        & $ K$  &  10.59 &  10.16 &   9.97 &   9.78 &   29.4 &   16.1 & N\nl
 2 & MCG-02-08-039 & SAB(rs)apec: & 0.0296 & $ I$  &  14.23 &  13.70 &  13.30 &  12.85 &   33.9 &   19.3 & E\nl
   &  &  &        & $ K$  &  11.86 &  11.37 &  11.00 &  10.65$^x$ &   43.8$^x$ &   17.4$^x$ & N\nl
 2 & MCG-03-58-007 & (R')SAB(s)0/a & 0.0315 & $ B$  &  15.62 &  15.22 &  14.98 &  14.96 &   17.7 &   19.2 & C92\nl
   &  &  &        & $ V$  &  14.79 &  14.39 &  14.16 &  14.11 &   19.6 &   19.3 & C92\nl
   &  &  &        & $ I$  &  13.68 &  13.29 &  13.07 &  13.03 &   19.4 &   18.2 & C92\nl
   &  &  &        & $ J$  &  12.47 &  12.13 &  11.94 &  11.91 &   18.9 &   17.0 & N\nl
   &  &  &        & $ H$  &  11.52 &  11.23 &  11.06 &  11.02 &   20.6 &   16.4 & N\nl
   &  &  &        & $ K$  &  10.94 &  10.73 &  10.62 &  10.61 &   18.1 &   16.1 & N\nl
 1 & MCG-03-07-011 & SAB(s)c & 0.0337 & $ V$  &  15.37 &  14.44 &  13.93 &  13.48 &   29.6 &   19.1 & L93\nl
 1 & MCG-05-13-017 & S0/a & 0.0126 & $ V$  &  13.98 &  13.54 &  13.31 &  13.14 &   29.6 &   19.7 & E\nl
   &  &  &        & $ J$  &  11.90 &  11.50 &  11.30 &  11.16 &   27.4 &   17.4 & I2\nl
   &  &  &        & $ H$  &  11.14 &  10.76 &  10.56 &  10.42 &   28.2 &   16.9 & I2\nl
   &  &  &        & $ K$  &  10.82 &  10.45 &  10.26 &  10.13 &   31.6 &   16.4 & I2\nl
 1 & MCG-06-30-015$^*$ & E-S0 & 0.0077 & $ B$  &  14.96 &  14.34 &  14.16 &  14.02 &   31.1 &   20.5 & C92\nl
   &  &  &        & $ V$  &  14.07 &  13.43 &  13.27 &  13.36$^x$ &   92.3$^x$ &   19.5$^x$ & C92\nl
   &  &  &        & $ I$  &  12.87 &  12.30 &  12.14 &  12.03 &   30.0 &   18.6 & C92\nl
 1 & MRK0006=IC0450 & SAB0+:. & 0.0185 & $ J$  &  11.86 &  11.47 &  11.29 &  10.86 &   54.2 &   19.4 & T\nl
   &  &  &        & $ H$  &  10.99 &  10.68 &  10.53 &  10.36 &   36.1 &   17.6 & T\nl
   &  &  &        & $ K$  &  10.38 &  10.14 &  10.03 &   9.90 &   37.8 &   17.7 & T\nl
 1 & MRK0009 & S0pec? & 0.0399 & $ V$  &  14.90 &  14.53 &  14.40 &  14.40 &   15.4 &   19.6 & L93\nl
   &  &  &        & $ K$  &  10.91 &  10.70 &  10.61 &  10.60 &   17.3 &   16.5 & LK\nl
 1 & MRK0079 & SBb & 0.0222 & $ V$  &  13.92 &  13.64 &  13.45 &  13.10 &   38.9 &   20.2 & L93\nl
   &  &  &        & $ K$  &  10.41 &  10.17 &  10.01 &   9.83 &   35.0 &   16.5 & LK\nl
 1 & MRK0231 & SA(rs)c?pec & 0.0422 & $ B$  &  14.73 &  14.52 &  14.39 &  14.32 &   26.5 &   21.4 & L90\nl
   &  &  &        & $ I$  &  12.40 &  12.17 &  12.04 &  11.71 &   46.0 &   20.4 & L90\nl
   &  &  &        & $ K$  &   9.02 &   8.92 &   8.88 &   8.82 &   32.4 &   17.2 & LK\nl
 2 & MRK0273$^*$ &  & 0.0378 & $ B$  &  15.74 &  15.20 &  15.02 &  14.81 &   37.5 &   22.3 & L90\nl
   &  &  &        & $ I$  &  14.30 &  13.79 &  13.61 &  13.36 &   40.2 &   21.1 & L90\nl
 1 & MRK0335 & S0/a & 0.0256 & $ V$  &  14.10 &  13.94 &  13.89 &  13.91 &   13.9 &   19.9 & L93\nl
   &  &  &        & $ J$  &  12.26 &  12.16 &  12.12 &  12.15 &   12.4 &   18.2 & N\nl
   &  &  &        & $ H$  &  11.14 &  11.07 &  11.04 &  11.05 &   13.7 &   17.9 & N\nl
   &  &  &        & $ K$  &  10.87 &  10.82 &  10.80 &  10.83 &   11.1 &   17.3 & N\nl
 2 & MRK0463 &  & 0.0497 & $ B$  &  15.81 &  15.13 &  14.93 &  15.29$^x$ &  102.9$^x$ &   20.6$^x$ & L90\nl
 1 & MRK0509 &  & 0.0344 & $ J$  &  11.87 &  11.73 &  11.65 &  11.61 &   20.5 &   18.6 & N\nl
   &  &  &        & $ H$  &  11.06 &  10.92 &  10.84 &  10.78 &   22.1 &   18.0 & N\nl
   &  &  &        & $ K$  &  10.16 &  10.06 &  10.01 &   9.96 &   23.1 &   17.8 & N\nl
 1 & MRK0618 & SB(s)bpec & 0.0355 & $ B$  &  13.36 &  12.81 &  12.59 &  12.40$^x$ &   52.2$^x$ &   20.2$^x$ & L90\nl
   &  &  &        & $ V$  &  14.81 &  14.35 &  14.12 &  14.04 &   23.8 &   18.7 & E\nl
   &  &  &        & $ J$  &  12.85 &  12.35 &  12.12 &  11.98 &   29.5 &   18.4 & I2\nl
   &  &  &        & $ H$  &  11.92 &  11.46 &  11.24 &  11.03 &   35.5 &   17.9 & I2\nl
   &  &  &        & $ K$  &  11.03 &  10.75 &  10.61 &  10.55 &   27.8 &   16.9 & I2\nl
 1 & MRK0704 & S & 0.0299 & $ V$  &  14.52 &  14.26 &  14.14 &  14.13 &   19.2 &   19.8 & L93\nl
   &  &  &        & $ K$  &  10.62 &  10.44 &  10.35 &  10.30 &   22.5 &   16.6 & LK\nl
\tablebreak
 1 & NGC4253=MRK766 & (R')SB(s)a: & 0.0129 & $ B$  &  14.65 &  14.13 &  13.87 &  13.55 &   35.7 &   20.3 & L90\nl
   &  &  &        & $ V$  &  14.11 &  13.62 &  13.37 &  13.09 &   31.1 &   19.3 & L93\nl
   &  &  &        & $ I$  &  13.41 &  12.87 &  12.59 &  12.30 &   29.9 &   18.5 & L90\nl
   &  &  &        & $ J$  &  11.97 &  11.51 &  11.28 &  11.04 &   29.6 &   16.8 & T\nl
   &  &  &        & $ H$  &  10.98 &  10.60 &  10.41 &  10.18 &   30.8 &   16.2 & T\nl
   &  &  &        & $ K$  &  10.10 &   9.86 &   9.72 &   9.56 &   30.6 &   16.1 & T\nl
 1 & MRK0817 & S? & 0.0315 & $ K$  &  10.49 &  10.27 &  10.16 &   9.90 &   31.9 &   18.1 & LK\nl
 2 & NGC0034=MRK0938 & Pec & 0.0198 & $ B$  &  14.96 &  14.32 &  14.10 &  13.96 &   26.7 &   19.9 & E\nl
   &  &  &        & $ V$  &  14.21 &  13.64 &  13.45 &  13.34 &   24.6 &   19.4 & E\nl
   &  &  &        & $ J$  &  11.88 &  11.55 &  11.42 &  11.34 &   25.8 &   18.4 & I2\nl
   &  &  &        & $ H$  &  10.95 &  10.70 &  10.61 &  10.56 &   22.3 &   17.2 & I2\nl
   &  &  &        & $ K$  &  10.46 &  10.24 &  10.17 &  10.13 &   20.5 &   16.8 & I2\nl
 1 & MRK1034 & S? & 0.0338 & $ V$  &  15.11 &  14.65 &  14.49 &  14.48 &   17.4 &   19.5 & L93\nl
   &  &  &        & $ J$  &  12.76 &  12.40 &  12.28 &  12.29 &   16.4 &   17.1 & N\nl
   &  &  &        & $ H$  &  11.99 &  11.63 &  11.51 &  11.50 &   17.1 &   16.4 & N\nl
   &  &  &        & $ K$  &  12.03 &  11.72 &  11.66 &  11.70 &   14.0 &   15.8 & N\nl
 1 & MRK1239 & E-S0 & 0.0199 & $ V$  &  14.82 &  14.43 &  14.36 &  14.38 &   13.2 &   18.4 & L93\nl
   &  &  &        & $ K$  &   9.54 &   9.44 &   9.41 &   9.40 &   18.8 &   16.8 & LK\nl
 2 & NGC0262=MRK0348 & SA(s)0/a: & 0.0151 & $ V$  &  14.88 &  14.29 &  14.00 &  13.15 &   53.8 &   21.8 & L93\nl
   &  &  &        & $ J$  &  12.52 &  12.01 &  11.76 &  11.26 &   39.2 &   19.1 & N\nl
   &  &  &        & $ H$  &  11.81 &  11.31 &  11.06 &  10.55 &   40.5 &   18.4 & N\nl
   &  &  &        & $ K$  &  11.57 &  11.12 &  10.89 &  10.56 &   30.5 &   17.9 & N\nl
 2 & NGC0424$^*$ & (R)SB(r)0/a & 0.0117 & $ B$  &  15.40 &  14.68 &  14.29 &  13.74$^x$ &   56.9$^x$ &   21.0$^x$ & E\nl
   &  &  &        & $ V$  &  14.49 &  13.78 &  13.39 &  12.88 &   53.9 &   20.1 & E\nl
   &  &  &        & $ I$  &  13.38 &  12.63 &  12.25 &  11.75 &   55.0 &   19.0 & E\nl
 2 & NGC0513 & S? & 0.0195 & $ V$  &  14.59 &  13.74 &  13.48 &  13.37 &   25.3 &   17.2 & L93\nl
   &  &  &        & $ J$  &  11.96 &  11.26 &  11.04 &  10.95 &   26.8 &   15.1 & N\nl
   &  &  &        & $ H$  &  11.45 &  10.77 &  10.56 &  10.49 &   25.6 &   14.3 & N\nl
   &  &  &        & $ K$  &  11.13 &  10.46 &  10.25 &  10.16 &   26.4 &   14.7 & N\nl
 1 & NGC0526A & S0pec? & 0.0192 & $ B$  &  15.85 &  15.40 &  15.19 &  14.79 &   31.8 &   22.0 & E\nl
   &  &  &        & $ J$  &  12.42 &  12.10 &  11.95 &  11.61 &   34.0 &   19.3 & I2\nl
   &  &  &        & $ H$  &  11.40 &  11.15 &  11.04 &  10.84 &   29.0 &   18.6 & I2\nl
   &  &  &        & $ K$  &  10.58 &  10.41 &  10.32 &  10.13 &   32.9 &   18.2 & I2\nl
 1 & NGC0931=MRK1040$^*$ & Sbc & 0.0167 & $ V$  &  14.83 &  14.18 &  13.81 &  12.86$^x$ &   82.5$^x$ &   20.9$^x$ & L93\nl
   &  &  &        & $ J$  &  12.27 &  11.64 &  11.30 &  10.65$^x$ &   77.0$^x$ &   18.4$^x$ & N\nl
   &  &  &        & $ H$  &  11.24 &  10.63 &  10.29 &   9.73$^x$ &   99.3$^x$ &   17.7$^x$ & N\nl
   &  &  &        & $ K$  &  10.69 &  10.15 &   9.85 &   9.28$^x$ &   85.0$^x$ &   17.1$^x$ & N\nl
 2 & NGC1056=MRK1183 & Sa: & 0.0052 & $ V$  &  14.16 &  13.58 &  13.30 &  12.80 &   45.8 &   20.4 & L93\nl
   &  &  &        & $ J$  &  11.33 &  10.79 &  10.56 &  10.17 &   47.2 &   17.8 & N\nl
   &  &  &        & $ H$  &  10.44 &   9.93 &   9.71 &   9.35 &   50.5 &   17.2 & N\nl
   &  &  &        & $ K$  &  10.12 &   9.62 &   9.41 &   9.10 &   50.0 &   17.0 & N\nl
 2 & NGC1068=M77 & (R)SA(rs)b & 0.0038 & $ V$  &  11.31 &  10.63 &  10.28 &   9.01 &  148.0 &   20.1 & L93\nl
 2 & NGC1125 & (R')SAB(rl:)0$^+$ & 0.0110 & $ V$  &  14.61 &  13.88 &  13.53 &  13.02 &   50.0 &   19.6 & L93\nl
 2 & NGC1143/44$^*$ & RingB & 0.0288 & $ B$  &  16.20 &  15.22 &  14.80 &  13.93 &   46.3 &   21.6 & E\nl
   &  &  &        & $ V$  &  14.98 &  14.12 &  13.72 &  12.79 &   48.0 &   20.5 & E\nl
   &  &  &        & $ J$  &  12.23 &  11.50 &  11.19 &  10.35 &   50.1 &   18.4 & I2\nl
   &  &  &        & $ H$  &  11.34 &  10.63 &  10.34 &   9.57$^x$ &   64.1$^x$ &   17.7$^x$ & I2\nl
   &  &  &        & $ K$  &  10.93 &  10.21 &   9.95 &   9.20 &   48.2 &   17.8 & I2\nl
 2 & NGC1194 & SA0+: & 0.0132 & $ V$  &  14.49 &  13.74 &  13.39 &  12.44 &   73.3 &   21.2 & L93\nl
   &  &  &        & $ J$  &  12.12 &  11.53 &  11.24 &  10.68 &   57.8 &   18.6 & N\nl
   &  &  &        & $ H$  &  11.31 &  10.74 &  10.46 &   9.95$^x$ &   63.0$^x$ &   18.0$^x$ & N\nl
   &  &  &        & $ K$  &  10.94 &  10.44 &  10.19 &   9.71 &   49.6 &   17.4 & N\nl
 2 & NGC1241 & SB(rs)b & 0.0135 & $ V$  &  14.55 &  13.85 &  13.41 &  12.23 &   80.0 &   20.3 & L93\nl
\tablebreak
 2 & NGC1320=MRK0607 & Sa:sp & 0.0091 & $ V$  &  14.33 &  13.64 &  13.33 &  12.81 &   57.2 &   19.8 & L93\nl
   &  &  &        & $ J$  &  11.63 &  11.09 &  10.83 &  10.42 &   43.4 &   17.4 & N\nl
   &  &  &        & $ H$  &  10.84 &  10.32 &  10.08 &   9.68 &   56.6 &   16.0 & N\nl
   &  &  &        & $ K$  &  10.45 &   9.97 &   9.74 &   9.38 &   53.0 &   15.7 & N\nl
 2 & NGC1667 & SAB(r)c & 0.0152 & $ B$  &  15.22 &  14.22 &  13.73 &  12.93 &   62.8 &   19.4 & E\nl
   &  &  &        & $ V$  &  14.23 &  13.37 &  12.92 &  12.21 &   45.6 &   18.4 & E\nl
   &  &  &        & $ K$  &  10.68 &   9.88 &   9.49 &   8.97 &   44.6 &   15.3 & I2\nl
 1 & NGC2639 & (R)SA(r)a:? & 0.0106 & $ V$  &  13.59 &  12.82 &  12.45 &  11.81 &   65.4 &   19.5 & L93\nl
   &  &  &        & $ J$  &  11.04 &  10.28 &   9.92 &   9.40 &   64.0 &   16.8 & X\nl
   &  &  &        & $ K$  &  10.04 &   9.27 &   8.91 &   8.44 &   63.2 &   16.0 & X\nl
 1 & NGC2992 & Sapec & 0.0077 & $ B$  &  15.31 &  14.50 &  14.02 &  13.07 &   79.9 &   21.8 & C92\nl
   &  &  &        & $ V$  &  14.25 &  13.48 &  13.04 &  12.21 &   80.5 &   20.9 & C92\nl
   &  &  &        & $ I$  &  12.76 &  12.05 &  11.68 &  10.85 &   72.7 &   19.9 & C92\nl
 2 & NGC3079 & SB(s)c & 0.0038 & $ V$  &  14.76 &  13.51 &  12.89 &  10.85 &  122.1 &   20.4 & L93\nl
 1 & NGC3227 & SAB(s)pec & 0.0039 & $ J$  &  10.90 &  10.42 &  10.10 &   8.99$^x$ &  125.0$^x$ &   17.1$^x$ & X\nl
   &  &  &        & $ K$  &   9.59 &   9.20 &   8.92 &   8.26$^x$ &  253.5$^x$ &   17.3$^x$ & X\nl
   &  &  &        & $ K$  &   9.46 &   9.07 &   8.81 &   7.24 &  142.1 &   17.0 & T\nl
 1 & NGC3516 & (R)SB(s)0$^0:$ & 0.0088 & $ K$  &   9.34 &   8.99 &   8.82 &   8.47 &   51.4 &   16.9 & T\nl
 1 & NGC4051 & SAB(rs)bc & 0.0024 & $ J$  &  11.22 &  10.73 &  10.42 &   8.75$^x$ &  208.8$^x$ &   18.3$^x$ & T\nl
   &  &  &        & $ H$  &  10.38 &   9.95 &   9.67 &   8.07 &  141.1 &   17.1 & T\nl
   &  &  &        & $ K$  &   9.98 &   9.61 &   9.36 &   7.69$^x$ &  207.7$^x$ &   17.5$^x$ & T\nl
 1 & NGC4151 & (R')SAB(rs)ab: & 0.0033 & $ V$  &  11.88 &  11.55 &  11.37 &  10.57$^x$ &  116.6$^x$ &   20.4$^x$ & L93\nl
   &  &  &        & $ K$  &   8.36 &   8.11 &   7.96 &   7.65$^x$ &   60.6$^x$ &   16.4$^x$ & LK\nl
 2 & NGC4388$^*$ & SA(s)b:sp & 0.0084 & $ V$  &  14.41 &  13.38 &  12.80 &  11.34$^x$ &  163.2$^x$ &   20.2$^x$ & CA\nl
   &  &  &        & $ I$  &  13.10 &  12.05 &  11.51 &  10.18$^x$ &  155.8$^x$ &   19.0$^x$ & CA\nl
 2 & NGC4501=M88 & SA(rs)b & 0.0076 & $ I$  &  11.68 &  10.83 &  10.37 &   8.57$^x$ &  350.9$^x$ &   19.1$^x$ & CA\nl
 1 & NGC4593 & (R)SB(rs)b & 0.0090 & $ V$  &  13.85 &  13.03 &  12.65 &  11.24 &  111.8 &   20.2 & L93\nl
   &  &  &        & $ K$  &  10.04 &   9.48 &   9.20 &   8.43$^x$ &   96.2$^x$ &   16.5$^x$ & LK\nl
 2 & NGC4968 & (R')SAB0$^0$ & 0.0099 & $ B$  &  15.80 &  15.13 &  14.83 &  14.17 &   55.5 &   21.8 & C92\nl
   &  &  &        & $ V$  &  14.79 &  14.15 &  13.86 &  13.24 &   54.1 &   20.9 & C92\nl
   &  &  &        & $ I$  &  13.52 &  12.90 &  12.62 &  11.96 &   56.2 &   19.8 & C92\nl
 1 & NGC5033 & SA(s)c & 0.0029 & $ K$  &   9.28 &   8.54 &   8.20 &   7.58$^x$ &   76.2$^x$ &   14.2$^x$ & LK\nl
 2 & NGC5135 & SB(l)ab & 0.0137 & $ B$  &  14.34 &  13.88 &  13.62 &  12.97 &   67.7 &   20.7 & C92\nl
   &  &  &        & $ V$  &  13.69 &  13.17 &  12.88 &  12.25 &   62.1 &   19.6 & C92\nl
   &  &  &        & $ I$  &  12.62 &  12.05 &  11.74 &  11.08 &   65.2 &   18.6 & C92\nl
 2 & NGC5256=MRK0266 & pec & 0.0279 & $ B$  &  15.53 &  14.61 &  14.19 &  13.85$^x$ &   40.7$^x$ &   20.9$^x$ & L90\nl
   &  &  &        & $ I$  &  14.21 &  13.33 &  12.93 &  12.52 &   35.0 &   19.8 & L90\nl
 2 & NGC5506$^*$ & Sapec & 0.0062 & $ B$  &  15.84 &  14.90 &  14.42 &  13.12 &   93.9 &   19.8 & C92\nl
   &  &  &        & $ V$  &  14.89 &  13.96 &  13.48 &  12.26 &   91.6 &   19.1 & C92\nl
   &  &  &        & $ I$  &  13.55 &  12.66 &  12.18 &  10.97 &   95.2 &   17.4 & C92\nl
   &  &  &        & $ K$  &   8.98 &   8.80 &   8.67 &   8.15 &   94.6 &   15.3 & T\nl
 1 & NGC5548 & (R')SA(s)0/a & 0.0172 & $ J$  &  11.73 &  11.26 &  11.02 &  10.58 &   41.8 &   18.4 & T\nl
   &  &  &        & $ K$  &  10.24 &   9.92 &   9.75 &   9.40 &   44.7 &   17.7 & T\nl
 2 & NGC5929$^*$ & Sab:pec & 0.0083 & $ V$  &  14.49 &  13.87 &  13.51 &  11.86$^x$ &   90.8$^x$ &   20.7$^x$ & CA\nl
   &  &  &        & $ I$  &  13.19 &  12.59 &  12.27 &  10.82 &   77.5 &   19.3 & CA\nl
   &  &  &        & $ J$  &  12.04 &  11.54 &  11.27 &   9.88 &   76.2 &   18.6 & N\nl
   &  &  &        & $ H$  &  11.29 &  10.80 &  10.53 &   9.16$^x$ &   84.4$^x$ &   17.9$^x$ & N\nl
   &  &  &        & $ K$  &  11.05 &  10.58 &  10.32 &   8.87$^x$ &   77.2$^x$ &   17.7$^x$ & N\nl
 2 & NGC5953 & SAa:pec & 0.0066 & $ I$  &  12.19 &  11.66 &  11.46 &  10.66 &   64.1 &   19.6 & CA\nl
   &  &  &        & $ J$  &  11.23 &  10.73 &  10.54 &   9.85 &   55.2 &   18.4 & N\nl
   &  &  &        & $ H$  &  10.39 &   9.91 &   9.73 &   9.08 &   58.0 &   17.7 & N\nl
   &  &  &        & $ K$  &  10.20 &   9.70 &   9.53 &   8.85 &   55.3 &   17.7 & N\nl
 2 & NGC6810 & SA(s)ab:sp & 0.0068 & $ B$  &  14.54 &  13.77 &  13.36 &  12.25 &   79.0 &   21.5 & C92\nl
   &  &  &        & $ V$  &  13.54 &  12.78 &  12.38 &  11.40 &   70.1 &   20.3 & C92\nl
   &  &  &        & $ I$  &  12.12 &  11.38 &  11.02 &  10.20 &   66.4 &   18.8 & C92\nl
 1 & NGC6860 & (R')SB(r)ab & 0.0149 & $ B$  &  14.85 &  14.25 &  13.94 &  13.64 &   36.6 &   19.6 & C92\nl
   &  &  &        & $ V$  &  14.05 &  13.45 &  13.15 &  12.86 &   35.8 &   19.2 & C92\nl
   &  &  &        & $ I$  &  12.98 &  12.39 &  12.11 &  11.86 &   34.4 &   18.0 & C92\nl
 2 & NGC6890$^*$ & (R')SA(r:)ab & 0.0081 & $ B$  &  15.18 &  14.23 &  13.75 &  13.06 &   48.5 &   19.7 & C92\nl
   &  &  &        & $ V$  &  14.26 &  13.38 &  12.94 &  12.28 &   46.9 &   18.9 & C92\nl
   &  &  &        & $ I$  &  13.13 &  12.29 &  11.86 &  11.26 &   45.3 &   17.6 & C92\nl
 2 & NGC7130=I5135$^*$ & Sapec & 0.0162 & $ B$  &  14.73 &  13.85 &  13.47 &  12.99 &   50.1 &   20.8 & C92\nl
   &  &  &        & $ V$  &  14.08 &  13.22 &  12.83 &  12.30 &   48.3 &   19.9 & C92\nl
   &  &  &        & $ I$  &  13.15 &  12.33 &  11.89 &  11.41 &   44.5 &   18.9 & C92\nl
   &  &  &        & $ J$  &  11.92 &  11.16 &  10.72 &  10.43$^x$ &   45.3$^x$ &   17.2$^x$ & I2\nl
   &  &  &        & $ H$  &  11.10 &  10.38 &   9.95 &   9.62$^x$ &   38.2$^x$ &   16.4$^x$ & I2\nl
   &  &  &        & $ K$  &  10.70 &  10.01 &   9.59 &   9.34$^x$ &   47.0$^x$ &   16.3$^x$ & I2\nl
 2 & NGC7172$^*$ & Sapec & 0.0087 & $ B$  &  15.60 &  14.77 &  14.27 &  12.80 &   85.2 &   20.9 & C92\nl
   &  &  &        & $ V$  &  14.59 &  13.76 &  13.27 &  12.01 &   70.8 &   19.2 & C92\nl
   &  &  &        & $ I$  &  13.01 &  12.20 &  11.75 &  10.71 &   69.0 &   17.4 & C92\nl
 1 & NGC7213 & SA(s)0$^0$ & 0.0060 & $ B$  &  13.51 &  12.74 &  12.29 &  11.07$^x$ &  124.7$^x$ &   21.2$^x$ & C92\nl
   &  &  &        & $ V$  &  12.58 &  11.82 &  11.38 &  10.16$^x$ &  123.5$^x$ &   20.3$^x$ & C92\nl
   &  &  &        & $ I$  &  11.42 &  10.67 &  10.26 &   9.08$^x$ &  120.7$^x$ &   19.1$^x$ & C92\nl
 1 & NGC7314$^*$ & SAB(rs)bc & 0.0047 & $ B$  &  15.86 &  14.78 &  14.14 &  11.78 &  128.2 &   18.9 & C92\nl
   &  &  &        & $ V$  &  14.96 &  13.90 &  13.29 &  11.10 &  126.1 &   18.8 & C92\nl
   &  &  &        & $ I$  &  13.82 &  12.78 &  12.20 &  10.19 &  120.5 &   17.9 & C92\nl
 1 & NGC7469 & (R')SAB(rs)a & 0.0163 & $ J$  &  10.93 &  10.60 &  10.42 &  10.05 &   48.2 &   18.6 & N\nl
   &  &  &        & $ H$  &  10.10 &   9.80 &   9.64 &   9.31 &   50.6 &   18.0 & N\nl
   &  &  &        & $ K$  &   9.60 &   9.39 &   9.28 &   9.14 &   35.3 &   17.0 & N\nl
 2 & NGC7496$^*$ & (R':)SB(rs)bc+SB(s)m & 0.0055 & $ B$  &  14.82 &  14.31 &  13.95 &  12.04 &  104.4 &   19.9 & C92\nl
   &  &  &        & $ V$  &  14.28 &  13.69 &  13.29 &  11.44 &  102.5 &   19.7 & C92\nl
   &  &  &        & $ I$  &  13.34 &  12.67 &  12.25 &  10.39 &  103.9 &   18.6 & C92\nl
 2 & NGC7582$^*$ & (R'$_1$)SB(s)ab & 0.0053 & $ B$  &  14.68 &  13.96 &  13.46 &  11.44 &  146.1 &   19.1 & C92\nl
   &  &  &        & $ V$  &  13.79 &  13.06 &  12.58 &  10.66 &  142.2 &   18.4 & C92\nl
   &  &  &        & $ I$  &  12.36 &  11.68 &  11.24 &   9.49 &  144.2 &   17.3 & C92\nl
 2 & NGC7590$^*$ & S(r?)bc & 0.0053 & $ B$  &  14.88 &  13.87 &  13.30 &  11.95 &   83.3 &   17.3 & C92\nl
   &  &  &        & $ V$  &  13.95 &  13.06 &  12.56 &  11.29 &   80.6 &   17.7 & C92\nl
   &  &  &        & $ I$  &  12.72 &  11.94 &  11.49 &  10.34 &   81.3 &   17.3 & C92\nl
 1 & NGC7603=MRK0530 & SA(rs)b:pec & 0.0295 & $ B$  &  15.15 &  14.72 &  14.45 &  13.55 &   59.7 &   22.2 & L90\nl
   &  &  &        & $ I$  &  14.15 &  13.61 &  13.31 &  12.91 &   35.5 &   20.0 & L90\nl
   &  &  &        & $ J$  &  11.85 &  11.39 &  11.15 &  10.73 &   44.6 &   18.1 & N\nl
   &  &  &        & $ H$  &  10.88 &  10.49 &  10.28 &   9.90 &   46.5 &   17.4 & N\nl
   &  &  &        & $ K$  &  10.20 &   9.91 &   9.75 &   9.40 &   43.2 &   17.2 & N\nl
 2 & NGC7674=MRK0533 & SA(r)bcpec & 0.0291 & $ J$  &  12.51 &  11.87 &  11.54 &  10.98 &   35.1 &   17.7 & N\nl
   &  &  &        & $ H$  &  11.67 &  11.07 &  10.76 &  10.17 &   36.1 &   16.9 & N\nl
   &  &  &        & $ K$  &  11.18 &  10.77 &  10.55 &  10.30 &   29.8 &   16.8 & N\nl
 2 & TOL1238-364=IC3639 & SB(rs)bc:S & 0.0110 & $ B$  &  14.65 &  13.99 &  13.60 &  13.07 &   37.5 &   18.4 & C92\nl
   &  &  &        & $ V$  &  13.92 &  13.31 &  12.97 &  12.52 &   34.9 &   18.0 & C92\nl
   &  &  &        & $ I$  &  12.98 &  12.40 &  12.10 &  11.70 &   34.0 &   17.9 & C92\nl
 2 & UGC11680=MRK0897 &  & 0.0263 & $ V$  &  15.13 &  14.86 &  14.73 &  14.75 &   39.4 &   22.7 & CA\nl
   &  &  &        & $ I$  &  14.01 &  13.68 &  13.51 &  13.36 &   51.0 &   22.0 & CA\nl
   &  &  &        & $ J$  &  12.79 &  12.55 &   0.00 &  11.82 &   43.2 &   20.9 & N\nl
   &  &  &        & $ H$  &  12.12 &  11.86 &  11.75 &  11.17 &   52.5 &   20.5 & N\nl
   &  &  &        & $ K$  &  11.90 &  11.69 &   0.00 &  11.62 &   35.7 &   19.8 & N\nl
\enddata 
\tablenotetext{*}{Sources for which $B$ and $V$ profiles extracted with
$I$-or $J$-band ellipse parameters.} 
\tablenotetext{\it x}{Isophotal radius is extrapolated, not interpolated.}
\end{deluxetable}